\pdfoutput=1
\documentclass[aps,twocolumn,preprintnumbers,longbibliography]{revtex4-1}
\usepackage{graphicx}
\usepackage{amsfonts}
\usepackage{amsmath}
\usepackage[svgnames]{xcolor}
\usepackage[colorlinks]{hyperref}
\hypersetup{linkcolor=DarkBlue,citecolor=DarkBlue,filecolor=black,urlcolor=DarkBlue}
\usepackage{multirow}
\usepackage{mathrsfs}
\usepackage{url}

\usepackage{amssymb}

\newcommand\Tr{\mathop{\mathrm{Tr}}}
\newcommand{\la}{\label}
\newcommand{\be}{\begin{equation}}
\newcommand{\ee}{\end{equation}}
\newcommand{\bea}{\begin{eqnarray}}
\newcommand{\eea}{\end{eqnarray}}
\newcommand{\p}{\partial}

\newcommand{\bx} {\mathbf{x}}
\newcommand{\bk} {\mathbf{k}}
\newcommand{\bq} {\mathbf{q}}
\newcommand{\mh} {\mathfrak{h}}

\newcommand{\comment}[1]{}

\usepackage{color}

\begin{document}

\title{Bimetric Theory of Fractional Quantum Hall States}

\author{Andrey Gromov}
\affiliation{Kadanoff Center for Theoretical Physics, University of Chicago, Chicago, Illinois 60637, USA}

\author{Dam Thanh Son}
\affiliation{Kadanoff Center for Theoretical Physics, University of Chicago, Chicago, Illinois 60637, USA}


\begin{abstract}

We present a bimetric low-energy effective theory of fractional quantum Hall (FQH) states that describes the topological properties and a gapped collective excitation, known as the Girvin-Macdonald-Platzman (GMP) mode. The theory consists of a topological Chern-Simons action, coupled to a symmetric rank-2 tensor, and an action \emph{\`a la} bimetric gravity, describing the gapped dynamics of a spin-$2$ mode. The theory is formulated in curved ambient space and is spatially covariant, which allows us to restrict the form of the effective action and the values of phenomenological coefficients. Using bimetric theory, we calculate the projected static structure factor up to the $k^6$ order in the momentum expansion. To provide further support for the theory, we derive the long-wave limit of the GMP algebra, the dispersion relation of the GMP mode, and the Hall viscosity of FQH states. The particle-hole (PH) transformation of the theory takes a very simple form, making the duality between FQH states and their PH conjugates manifest. We also comment on the possible applications to fractional Chern insulators, where closely related structures arise. It is shown that the familiar FQH observables acquire a curious geometric interpretation within the bimetric formalism.

\end{abstract}

\maketitle

\newpage

\section{Introduction}

   During the last two decades, topological quantum field theory (TQFT) has firmly established itself as a useful low-energy theory of fractional quantum Hall states \cite{Tsui:1982yy, laughlin1983anomalous} (and, more generally, of topological phases in two spatial dimensions). It describes the properties of local anyonic quasiparticles and the ground-state degeneracy on higher-genus surfaces; in addition, it allows one to calculate the universal part of the linear response to external fields and implies gapless edge excitations. TQFT was introduced into quantum Hall physics in the seminal papers \cite{ZHK1989, read1989order} and has led to a satisfactory picture of the fractionalized local excitations originally introduced by Laughlin in the language of trial wave functions \cite{laughlin1983anomalous}. The relationship between the two approaches was first explained in Ref.~\cite{moore1991nonabelions}. The properties described by the TQFT are expected to be insensitive to the microscopic details of the material or the experimental setup that probes the topological phase in question.
   
  In recent years, it was realized that additional universal features are revealed when a quantum Hall state (or any topological phase in two dimensions) is placed on a curved surface. Paradoxically, when the TQFT is coupled to the geometry of the ambient space, it produces nontrivial linear response functions that encode additional information about a quantum Hall state such as the Wen-Zee shift \cite{haldane1983fractional, WenZeeShiftPaper, 1993-frohlich}, the Hall viscosity \cite{avron1995viscosity, tokatly2007lorentz, read2009non, haldane2009hall}, and the central charge \cite{Abanov-2014, klevtsov2014random, ferrari2014fqhe, CLW, GCYFA, bradlyn2015topological, klevtsov2015precise}.  

 Another remarkable feature of fractional quantum Hall states, not shared generically by other topological phases of matter, is the presence of a gapped collective excitation first proposed by Girvin, Macdonald, and Platzman (GMP) \cite{girvin1986magneto}. This excitation is absent in the integer quantum Hall phases but appears to be universally present in the fractional states and has been experimentally observed \cite{Pinczuk, Pinczuk2, kukushkin2009dispersion}. One remarkable property of the GMP mode is that it carries angular momentum $L=2$ at zero linear momentum.
This property of the GMP mode is one of the motivations that lead Haldane to propose that fractional quantum Hall states have a hidden sector described by a gapped effective theory of a geometric nature \cite{haldane2009hall, haldane2011geometrical, haldane2011self}.  More recently, the authors of Refs.~\cite{maciejko2013field, you2014theory} (see also Ref. \cite{regnault2016evidence}), motivated by recent experiments \cite{xia2011evidence}, suggested that the GMP mode can be understood as gapped fluctuations of a nematic order parameter; hence, it becomes light near the nematic phase transition. This interpretation justifies the inclusion of only the spin-2 mode in the effective field theory. However, it leaves one wondering about the role of general covariance and geometry emphasized in Refs.~\cite{haldane2009hall, haldane2011geometrical, haldane2011self}. Another effective theory of the GMP mode was considered in Ref.~\cite{golkar2016spectral}, where the Wess-Zumino-Witten action was used to construct the kinetic term. Geometric degrees of freedom and possible means of their observation were also discussed in Refs.~\cite{yang2013geometry, yang2016acoustic} in a different language. 
  
 The goal of the present paper is to construct an effective theory describing the spin-2 mode that is consistent with all constraints arising from the topological properties and the structure of the single Landau level. Our approach combines the ideas of Refs.~\cite{haldane2009hall, haldane2011geometrical, haldane2011self, maciejko2013field, you2014theory} with the formalism developed in Ref.~\cite{AGBB} and the bimetric approach to massive gravity \cite{bimetric2010, bergshoeff2013zwei}. This effective theory is geometric in nature, covariant with respect to spatial diffeomorphisms, and nonlinear by construction. Similarly to Ref.~\cite{maciejko2013field}, the theory consists of a Chern-Simons TQFT interacting with a massive spin-2 field, whose dynamics is governed by an action reminiscent of the action for bimetric theory of massive gravity. We evaluate the projected static structure factor (SSF), which is given by the equal-time correlation function of the Ricci scalar within bimetric theory, and we match it to the microscopic result of Ref.~\cite{girvin1986magneto}. This matching completely fixes all dimensionless phenomenological parameters in the theory and provides a matching of dimensionful parameters to quantities computable from the microscopics. Then, we use the theory to rederive the long-wavelength limit of the GMP algebra, the dispersion relation of the GMP mode, and the Hall viscosity, and we show that all linear response functions are reproduced correctly.  The proposed theory is valid as long as the observable quantities are saturated by the single-mode approximation (SMA). This is presumably true close to a nematic phase transition.  While there is \emph{a priori} no reason to expect the SMA to be reliable away from a nematic phase transition, there is evidence that it is a good approximation numerically \cite{platzman1996resonant, Papic-SMA, jolicoeur2017shape}.
 
 There are two equivalent formulations of the bimetric theory. In the first formulation, the metric sets the background geometry, and the dynamical degree of freedom is a $2\times 2$ symmetric matrix of unit determinant, reminiscent of a Goldstone field, that transforms under the internal symmetries of the tangent space. This matrix is not a metric and, in general, cannot be used to measure distances. This formulation is very close in spirit to the one in Ref.~\cite{maciejko2013field}.  In the second formulation, the quantum degree of freedom is a dynamical metric $\hat g_{ij}$. The two formulations are related to each other by a linear change of variables. We refer to these formulations as first and second order, correspondingly. 
 We emphasize that the presented theory does not refer to the notion (or any properties of) holomorphic wave functions and should be applicable for {\it any} Landau level, as well as fractional Chern insulators (with the possible incorporation of lattice symmetries).
 
 The plan of the paper is as follows. In Sec.\ II we review the bimetric geometry of Ref.~\cite{AGBB}. In Sec.\ III we introduce the bimetric effective theory, fix the phenomenological coefficients by matching the SSF computation, and reproduce the long-wave limit of the GMP algebra. In Sec.\ IV, we further develop the theory and discuss the role of the gravitational Chern-Simons term. In Sect.\ V, we present our conclusions and discuss open questions.

\section{Bimetric geometry}

In this section, we review the geometric formalism recently developed in Ref.~\cite{AGBB} to analyze anisotropic FQH states. This formalism will play the central role in formulating the effective theory.

\subsection{``Nonrelativistic'' geometry}

We start with a brief review of the geometry used to probe FQH states \cite{2012-HoyosSon, son2013newton, Abanov-2014, bradlyn2014low, gromov-thermal, Jensen2014coupling, gromov2016boundary}. In what follows, we do not assume any nongeneric symmetry such as rotational, Galilean, or Lorentz invariance, which will be reflected in the geometry we discuss. We describe the geometry using vielbein fields $e^A_\mu = (e^A_0, e^A_i)$ and $E^\mu_A =(E^0_A, E^i_A)$, where the indices $i,j, \ldots=1,2$ label the spatial coordinates on the manifold, while $A,B, \ldots=1,2$ are the internal indices. Here, $E^i_A$ is the inverse matrix of $e^A_i$.
Objects carrying the index $i$ transform under spatial coordinate transformations, and those carrying the index $A$ transform under internal $SO(2)$ rotations. Greek letters $\mu,\nu, \ldots=0,1,2$  are used for spacetime indices; however, we only allow time-indepedent spatial coordinate transformations, thus separating time from space. The spatial metric is then given by
\be\la{eq:metric}
g_{ij} = \delta_{AB} e^A_i e^B_j\,, \qquad g^{ij} = \delta^{AB} E_A^i E_B^j\,.
\ee 
We allow the metric to depend on both time $t$ and space $\mathbf{x}$. The spatial metric is used to measure spatial distances according to $ds^2 = g_{ij}(t,\mathbf{x}) dx^i dx^j$; thus, we allow the distance between any two points to change in time. The $SO(2)$ ``symmetry'' that acts on the vielbein field merely reflects the inherent ambiguity of splitting the metric into a product of two matrices. Any physical observable must not depend on how this ambiguity is resolved, which will translate into the invariance of the effective action and the generating functional with respect to local $SO(2)$ transformations.

We introduce a covariant derivative $\nabla_\mu$ and impose metric compatibility conditions:
\be\la{eq:vielbeinpost}
\nabla_\mu e^A_\nu =0\,,\qquad \nabla_\mu g_{ij} =0\,.
\ee
Defining the action of the covariant derivative to be
\be
\nabla_\mu e^A_\nu = \p_\mu e^A_\nu - \Gamma^\lambda{}_{\nu,\mu} e^A_\lambda + \omega^A{}_{B,\mu} e^B_\nu = 0\,,
\ee
we find that the spin connection is given in terms of the vielbeins and the Christoffel symbols,
\bea
\omega_0 &=& \frac{1}{2} \epsilon_A{}^B E^i_B \p_0 e_i^A\,, \la{eq:omega0}   \\ 
\omega_j &=& \frac{1}{2} \epsilon_A{}^B \Big( E^i_B \p_j e_i^A - \Gamma^k{}_{i,j} e^A_k E_B^i\Big)\, \la{eq:omegaj}\,,
\eea
where the Christoffel symbols are determined from the second condition in Eq.\ \eqref{eq:vielbeinpost},
\bea
\Gamma^{i}{}_{k,j}&=&\frac{1}{2}g^{i\ell}\left(\partial_jg_{k\ell}+\partial_kg_{j\ell}-\partial_\ell g_{jk}\right)\label{eq:christoffel1}\,,\\
\Gamma^i{}_{j,0}&=&\frac{1}{2}g^{ik}\partial_0 g_{jk}\label{eq:christoffel2}\,.
\eea

Certain components of the Christoffel connection remain undetermined by Eq.~\eqref{eq:vielbeinpost} and, therefore, have to be determined solely by a torsion. We set the ``reduced torsion'' of Ref.~\cite{bradlyn2014low} to $0$. It is easy to verify that under a rotation of vielbeins, the spin connection $\omega_\mu$ transforms like an Abelian gauge field. The Ricci curvature of a time slice is given by
\be
R= \frac{2}{\sqrt{g}} \Big(\partial_1 \omega_2 - \partial_2 \omega_1\Big)\,.
\ee
The Ricci curvature can depend on time; however, we assume that the Euler characteristic
\be
\chi = \frac{1}{4\pi} \!\int\! \sqrt{g} R
\ee
is time independent. Finally, we often use the form notation $d\omega$, which means
\be
(d\omega)_{\mu\nu} = \p_\mu \omega_\nu - \p_\nu \omega_\mu\,.
\ee 
This completes our review of the spatial geometry. In the remainder of the paper, this geometry will describe the shape and the curvature of the sample where the topological electron fluid ``lives.''

\subsection{Intrinsic geometry}

Next, we recall the formalism developed in Ref.~\cite{AGBB} in the context of anisotropic FQH states. In the present context, there is no anisotropy in the sense of Ref.~\cite{AGBB}; however, we postulate the existence of a rank-$2$ symmetric tensor, $\mathfrak h_{AB}(\bx,t)$ that describes the spin-$2$ massive collective excitation. Note that, by definition, $\mathfrak h_{AB}(\bx,t)$ is a spacetime scalar that transforms only under the internal $SO(2)$. With this tensor at hand, we can reconstruct the formalism of Ref.~\cite{AGBB}.

Following Ref.~\cite{AGBB}, we introduce an analogue of the vielbeins according to
\be\la{eq:mh}
\mathfrak h_{AB} = \lambda_A^\alpha \lambda_B^\beta \delta_{\alpha \beta}\,,
\ee
where the indices $\alpha,\beta,\ldots =1,2$ correspond to a new ambiguity in splitting $\mathfrak h_{AB}$ into a product of two matrices. We denote this ambiguity as $\widehat{SO}(2)$. We use a convention in which all internal indices $A,B,\ldots$ and $\alpha,\beta,\ldots$ are raised and lowered with $\delta_{AB}$ and $\delta_{\alpha\beta}$, respectively.
 We also introduce the inverse of $\mathfrak h_{AB}$ which is denoted $\mathfrak H^{AB}$ and is, in general, \emph{not} equal to $\mathfrak h^{AB}$. Correspondingly, we introduce the inverse vielbeins:
\be
\mathfrak H^{AB} = \Lambda^A_\alpha \Lambda^B_\beta \delta^{\alpha\beta}\,.
\ee
Then,
\be
\mathfrak h_{AB} \mathfrak H^{BC} = \delta^B_C\,,\qquad \Lambda^A_\alpha  \lambda_A^\beta = \delta^\beta_\alpha\,.
\ee
Given these data, we can introduce an intrinsic vielbein field $\hat e^\alpha_\mu$ and an intrinsic metric $\hat g_{ij}$ according to
\be
\hat e^\alpha_\mu = e^A_\mu \lambda_A^\alpha\,,\qquad \hat g_{ij} = \hat e^\alpha_i  \hat e^\beta_j \delta_{\alpha\beta} = \mathfrak h_{AB} e^A_i e^B_j\,.
\ee
Next, we introduce the inverse intrinsic metric
\be
\hat G^{ij}  = \mathfrak H^{AB} E^i_A E^j_B = \hat E^i_\alpha \hat E^j_\beta \delta^{\alpha\beta}\,,
\ee
where we have also introduced the inverse second vielbein $\hat E^\mu_\alpha = E^\mu_A \lambda^A_\alpha$. We use a convention where the spatial indices $i,j,\ldots$ are raised and lowered with the spatial metric $g_{ij}$. 

With $\hat e^\alpha_\mu$ at hand, we again define a covariant derivative $\hat \nabla_\mu$ and impose the compatibility with the vielbein
\be
\hat \nabla_\mu \hat e^A_\nu = \p_\mu \hat e^A_\nu - \hat \Gamma^\lambda{}_{\nu,\mu} \hat e^A_\lambda + \hat \omega^A{}_{B,\mu} \hat e^B_\nu = 0\,,
\ee
which will again define the spin connection $\hat \omega_\mu$
\bea
\hat \omega_0 &=& \frac{1}{2} \epsilon_\alpha{}^\beta \hat E^i_\beta \p_0 \hat e_i^\alpha\,, \la{eq:omegahat0}   \\ 
\hat \omega_j &=& \frac{1}{2} \epsilon_\alpha{}^\beta \Big( \hat E^i_\beta \p_j \hat e_i^\alpha - \hat\Gamma^k{}_{i,j} \hat e^\alpha_k \hat E_\beta^i\Big)\,, \la{eq:omegahatj}
\eea
where the second Christoffel connection is defined by the condition $\hat \nabla_\mu \hat g_{ij} =0$ and is given by
\bea
\hat\Gamma^{i}{}_{k,j}&=&\frac{1}{2}\hat G^{i\ell}\left(\partial_j \hat g_{k\ell}+\partial_k\hat g_{j\ell}-\partial_\ell \hat g_{jk}\right)\,,\label{eq:christoffelhat1}\\
\hat\Gamma^i{}_{j,0}&=&\frac{1}{2}\hat G^{ik}\partial_0 \hat g_{jk}\,.\label{eq:christoffelhat2}
\eea
Clearly, the spin connection $\hat \omega_\mu$ transforms as an Abelian gauge field under the $\widehat{SO}(2)$ transformations. Next, we define the Ricci scalar according to $\hat R = \frac{2}{\sqrt{g}} \Big(\p_1 \hat\omega_2 - \p_2\hat\omega_1\Big)$ and the second Euler characteristic $\hat \chi$,
\be
\hat \chi = \frac{1}{4\pi} \!\int\! \sqrt{\hat g} \hat R\,.
\ee
It is not hard to see that
\be
\hat \chi = \chi + \mathcal X\,,
\ee
where $\chi$ is the Euler character of the physical space and $\mathcal X$ is the number of singularities in $\lambda^\alpha_A$. The latter can be evaluated as
\be
\mathcal X = \frac{1}{4\pi} \!\int\! \sqrt{\mathfrak h} \hat R\Big|_{g_{ij} = \delta_{ij}}\,,
\ee
where we have introduced $\mathfrak h = \lambda^2 = \det \mathfrak h_{AB} = (\det \lambda_A^\alpha)^2$.

With two independent metrics and connections, we can introduce extra data absent in the traditional Riemannian geometry. Consider the one-form
\be\la{eq-CDEF}
C^i{}_{j,\mu} = \Gamma^i{}_{j,\mu} - \hat \Gamma^i{}_{j,\mu}\,.
\ee
As a difference of two connections, $C^i{}_{j,k}$ transforms like a rank-$3$ tensor.  There are no more independent objects of interest. One can construct two independent one-forms from $C^i{}_{j,k}$: the trace $C^i{}_{i,k} \sim \p \ln  \mh $ and the antisymmetric part $C_\mu = \epsilon_i{}^jC^i{}_{j,\mu}$.  The latter does not vanish in our setup and will be used in the effective action.  Note that $C_\mu$ has good transformation properties only when the same diffeomorphism is applied simultaneously to $\Gamma$ and $\hat \Gamma$. Thus, any action that involves $C_\mu$ will break the two copies of diffeomorphisms (acting on $g$ and $\hat g$, correspondingly) down to a diagonal subgroup.

While the general geometric structure allows for arbitrary nondegenerate, positive-definite $\mathfrak h_{AB}$, we also impose a constraint 
\be\la{eq:const1}
\det \mathfrak h_{AB} =1\,,
\ee
which prohibits ``dilaton'' excitations of $\hat g_{ij}$ through the constraint
\be\la{eq:const2}
\det \hat g_{ij} = \det g_{ij}\,.
\ee
The  intrinsic metric will be viewed as a \emph{dynamical} property of the physical system. We can use either the internal field $\lambda^\alpha_B$ or the vielbein $\hat e_i^\alpha = e_i^B \lambda^\alpha_B$ to describe the physical degrees of freedom. These two choices correspond to the first- and second-order formalisms alluded to in the Introduction and are related by a linear, nondegenerate change of variables. Bimetric theory will be constructed in the second-order formalism since the symmetries are more transparent this way. In the second-order formulation, we impose the constraint \eqref{eq:const2} without any reference to $\mh_{AB}$. Finally, we note that a pair of metrics is precisely the starting point for building bimetric theory of massive gravity \cite{bimetric2010, de2014massive}, with the difference that  the theory there is Lorentz invariant and all of the components of both metrics are allowed to fluctuate.

Before moving on, we make a comment about the relation to the work of Haldane \cite{haldane2009hall, haldane2011geometrical, haldane2011self}. The geometric description discussed in Refs.\ \cite{haldane2009hall, haldane2011geometrical, haldane2011self} is represented by a unimodular (i.e., unit-determinant) metric, which, in our notation, is $\hat g_{ij}$. To avoid confusion and to impose the constraints coming from self-consistency in weakly curved space, the effective theory will be constructed when both the ambient and the ``dynamic''  spaces are curved from the very beginning.

\section{Bimetric effective theory}

In this section, we construct a covariant bimetric effective action for an Abelian FQH state that will include the massive dynamics of the GMP mode. Using this theory, we calculate the Hall viscosity and the projected static structure factor, and derive the long-wave limit of the GMP algebra.

We start with the more familiar topological part of the effective action, which includes the coupling of an internal gauge field $a$ to the bimetric geometry through a Wen-Zee term \footnote{We have explicitly assumed that Chern-Simons fields do not couple to the one-form $C$. Such coupling may be included and is discussed in Appendix \ref{app:linearized}.} 
\be\la{eq:Stop}
S_{\rm top} = \frac{k}{4\pi} \!\int\! ada - \frac{1}{2\pi} \!\int\! a d A - \frac{s}{2\pi}\!\int\! a d \omega - \frac{\varsigma}{2\pi}\!\int\!  a d \hat\omega\,,
\ee
where $k$ determines the filling factor $\nu = \frac{1}{k}$, whereas $s$ and $\varsigma$ describe the coupling to the ambient and dynamic geometries, respectively. The topological effective action describes local anyonic quasiparticles and their fractional electric charge, spin, and statistics. Integrating out the gauge field $a$ with a proper gauge-fixing condition \cite{GCYFA} leads to a generating functional $W[A,\omega,\hat \omega]$, which describes the kinematics of the metric $\hat g$ and the linear response functions---Hall conductance, Hall viscosity, and the shift. 

The electron density is given by
\be\la{eq:density}
\rho = \frac{\nu}{2\pi} B + \frac{\nu s}{4\pi} R + \frac{\nu \varsigma}{4\pi} \hat R\,,
\ee
and the total number of states is given by
\be\la{eq:shift2}
N = \nu N_\phi + \nu s \chi + \nu \varsigma  \chi\,,
\ee
where we have introduced the number of flux quanta $N_\phi = \frac1{2\pi} \!\int\! B$ and assumed that, as long as the fluctuations of $\hat g_{ij}$ are small, the two Euler characteristics are equal.

To interpret the meaning of Eq.\eqref{eq:shift2}, we recall the topological quantum number known as the shift \cite{haldane1983fractional, WenZeeShiftPaper, 1993-frohlich}, defined as an offset between the number of electrons and the number of flux quanta on a compact Riemann surface with the Euler characteristic $\chi$ according to
\be\la{eq:shiftdef}
N = \nu N_\phi + \nu \mathcal S \frac{\chi}{2}\,.
\ee
As a topological quantum number, the shift cannot change continuously when a small, translationally invariant perturbation is introduced. The shift is well defined and quantized even when the global rotational invariance is absent, as long as translational invariance is preserved \cite{AGBB}. The shift is used to distinguish topologically different FQH states that occur at the same filling factor. It is readily available in numerics and is often measured on a sphere.

Comparing Eq.\eqref{eq:shift2} with the definition \eqref{eq:shiftdef}, we find that the shift is given by
\be\la{shift}
\mathcal S  = 2(s+\varsigma) \,.
\ee
The electric current is given by
\be
j^i = \frac{\nu}{2\pi} \epsilon^{ik} E_k + \frac{\nu s}{2\pi} \epsilon^{ik} \mathcal E_k + \frac{\nu \varsigma}{2\pi} \epsilon^{ik} \hat{\mathcal E}_k\,,
\ee
where $\mathcal E_i =- \p_i \omega_0 + \p_0 \omega_i$ is the geometric analogue of the electric field. The conservation of the electric charge (in the absence of external fields) holds identically,
\be
\p_0 \rho + \p_i j^i = \p_0 \hat R +\epsilon^{ik}  \p_i \hat{\mathcal E}_k \equiv 0\,.
\ee

Finally, we note that the topological part of the effective action, $S_{\rm top}$, has an enlarged symmetry group. It can be directly verified that, in the absence of the electric field, $S_{\rm top}$ is invariant under a time-independent $SL(2,\mathbb R)$ transformation $\hat e^\alpha_i \rightarrow U_{\alpha}{}^\beta \hat e^\alpha_i$. This explains the appearance of the $\mathfrak{sl}(2,\mathbb R)$ Lie algebra, as we discuss later.

Next, we construct the dynamical part of the effective action. In fact, the parity-breaking terms will be generated after integrating out the gauge field $a$, and the parity-even terms will be added by hand. First, we need a kinetic term for $\mathfrak h_{AB}$. The obvious choice would be the Einstein-Hilbert action
\be
S_{EH} = \int\! d^3 x\, \sqrt{\hat g} \hat R\,;
\ee
however, this action is equal to the Euler characteristic (integrated over time) and does not generate any dynamics. 

It turns out that it is possible to construct another kinetic term, using $C_{k} = \epsilon^{j}{}_i C^i{}_{j,k}$, defined in Eq.\eqref{eq-CDEF}. Indeed, consider the term
\begin{multline}\la{eq:Skin}
S_{\rm kin}[\hat g;g] = - \frac{\alpha}{4}\! \int\! d^3x\, \sqrt{g}~  g^{kl} C_{k} C_{l} \\ \sim -\frac{\alpha}{4} \!\int\! d^3 x\, \sqrt{g} \left| \Gamma - \hat \Gamma \right|^2\,.
\end{multline}
This term is the same order in derivatives as the Einstein-Hilbert action and is diffeomorphism invariant. In the first-order language, as we will see shortly, $S_{\rm kin}[\hat g;g]$ is a covariant version of the ordinary kinetic term $\p \lambda \p \Lambda$.

 In the absence of higher-derivative terms, the coefficient $\alpha$ has to be strictly positive in order for the theory to be stable. It is illuminating to view $S_{\rm kin}[\hat g;g]$ as a functional of the Christoffel connection $\hat \Gamma$. From the perspective of the connection, $S_{\rm kin}$ is a potential term (since it is a polynomial in $\hat\Gamma$) that favors the configurations where the two connections are equal to each other: $\Gamma = \hat \Gamma$. This can be achieved if the two metrics are equal: $g_{ij} = \hat g_{ij}$. We emphasize here that  were we not careful about diffeomorphism invariance, we could have written a kinetic term $|\Gamma - \kappa \hat \Gamma|^2$, which would allow for solutions $\hat g_{ij} = \frac{1}{\kappa}  g_{ij}$ (reminiscent of $a = \frac{1}{k}A$ in Chern-Simons theory). However, these solutions, within our construction, are inconsistent with diffeomorphism invariance.

The solution we have just described is not the only one. An arbitrary solution is of the from $\hat g_{ij} = g_{ij}+h_{ij}$, where $h_{ij}$ is a \emph{constant} rank-2 tensor. Such a solution is acceptable since the Christoffel symbol behaves as $\hat \Gamma_{k} \sim \epsilon^{ij}\p_i \hat g_{jk}$. The condition $\det g = \det \hat g$ must be preserved. These solutions will parametrize the space of ground states. One way to fix $h_{ij}$ would be to choose a particular boundary condition for the metric $\hat g_{ij}$ at spatial infinity. Alternatively, we could choose a potential term that energetically favors a particular choice of $h_{ij}\neq 0 $. Such a potential will induce  a spontaneous breaking of the rotational symmetry (in flat space) since a symmetric matrix $\hat g_{ij}= \delta_{ij} + h_{ij} \neq \delta_{ij} $ is not an $SO(2)$ invariant tensor \footnote{In curved space, the presence of constant $h_{ij}$ leads to a breakdown of diffeomorphism invariance}. Massless fluctuations of $\hat g_{ij}$ around $h_{ij}$ will describe the Goldstone mode in the nematic (i.e., symmetry-broken) phase. We will elaborate on this phase in a separate publication.  Thus, we have established that $S_{\rm kin}$ forces the dynamical metric $\hat g_{ij}$ to follow the fixed spatial metric $g_{ij}$, and, consequently, it forces (in the isotropic phase) $\mh_{AB}=\delta_{AB}$.
 
If $\alpha<0$, the theory would favor the metric that deviates arbitrarily far from the fixed spatial metric, which leads to an instability. This instability will be seen in the dispersion relation of the GMP mode that goes to arbitrarily small negative energies at higher momentum, unless stabilized by higher-order terms in the kinetic-energy part of the effective action. In the second-order language, the interpretation of the instability is also clear---large values of $\hat \Gamma$ will be energetically favorable, but since the Christoffel connection $\hat \Gamma \sim \p \hat g$, the only way to maximize it is to make a configuration of $\hat g$ with very rapid spatial variations. The dynamical curvature $\hat R$ generated in such a way will be singular everywhere.  To describe the GMP mode, we have to enter this unstable regime; however, we argue that this instability can be cured by higher-order terms in $S_{\rm kin}$. Indeed, it is possible to add higher-derivative corrections to $S_{\rm kin}$ with a Lagrangian of the form
\be
\delta S_{\rm kin}^{(n)}[\hat g;g] =  \frac{\alpha_n}{4} \!\int\! d^3x\, \sqrt{g} g^{kl}  C_{k} (g^{ij}D_i D_j)^n C_{l}\,,
\ee
  where $D_i$ is the covariant derivative. The coefficients $\alpha^{(n)}$ can {\it a priori} be either positive or negative. As we show, choosing a positive $\alpha^{(2)}\approx |\alpha|$ will induce a roton minimum that will fix the instability. There are other terms that can contribute to the $(k\ell)^4$ power in the dispersion relation. Notably, the gravitational analogue of the ``Maxwell-type'' term is $\mathscr L \sim  c_1|\hat{\mathcal E}|^2 + c_2 \hat R^2$.  These terms describe the local current-current and density-density interactions, correspondingly. We do not consider higher-gradient terms in much detail but only point out that there is a variety of terms that can be added to the effective action to stabilize the GMP mode.

Next, we need to introduce a potential term that will force our spin-$2$ degree of freedom to be gapped. The choice of the potential is the defining feature of the standard bimetric gravity \cite{bimetric2010}, where the potential is carefully crafted to ensure the absence of ghosts. We discuss these potentials and their implications for our theory in Appendix \ref{app:potential}. Here, we only mention that, unlike the potential discussed below, the bimetric gravity potentials do not support a nematic phase transition.

We choose a potential equivalent to the potential in Ref.~\cite{maciejko2013field}. The potential term is given by
\be\la{eq:Spot}
S_{\rm pot}[\hat g;g] = -\frac{\tilde m}{2} \!\int\! d^3x\, \sqrt{g} \left(\frac{1}{2} \hat g_{ij} g^{ij} - \gamma  \right)^2\,,
\ee 
where we assumed $\tilde m>0$. Such a term is possible because of the bimetric nature of the theory.

 To understand the phase diagram, we note that $\hat g_{ij} g^{ij} = \Tr \mh$. Since $\mh$ is a symmetric unimodular matrix with positive eigenvalues, its trace satisfies $\Tr \mh \geq 2$. Thus, this potential supports two phases connected by a phase transition occurring at $\gamma = 1$. When $\gamma >1$, the system is in the symmetry-broken phase with $\Tr \mathfrak h = 2\gamma>2$ (and $h_{ij}\neq0$), and when $\gamma<1$, the system is in the isotropic phase with $\Tr \mathfrak h = 2$, $\mh_{AB} = \delta_{AB}$, and $\hat g_{ij} = g_{ij}$. Note that, by construction, in either phase the configuration of $\hat g_{ij}$ that minimizes $S_{\rm pot}[\hat g;g]$ also extremizes $S_{\rm kin}[\hat g;g]$.  Finally, we note that potential and kinetic terms  depend on the ambient metric $g_{ij}$, while the topological terms do not.

An object similar to $\hat g_{ij}$ has appeared in the study of fractional Chern insulators (FCI) \cite{roy2014band, jackson2015geometric}, where it was referred to as the ``quantum metric.'' The ``metric'' $g^{\alpha}_{\mu\nu}$ of Ref.~\cite{jackson2015geometric} is related to ours via $\hat g_{ij} = \frac{2}{B^\alpha} g^{\alpha}_{ij}$, where we have changed the type of indices to fit our notations, and $B^\alpha$ is the Berry curvature. Interestingly, Ref.~\cite{jackson2015geometric} investigated the dependence of the spectral gap on the Brillouin-zone-averaged difference $\langle T \rangle =\langle \Tr \mh \rangle -2 $. In our theory, the gapped phase occurs at $\langle T \rangle=0$, and the gapless phase appears at $\langle T\rangle >0$.  Thus, $\langle T\rangle$ is an order parameter for the nematic transition. In Ref.~\cite{jackson2015geometric},  the gap is observed at positive $\langle T\rangle$, and the gap decreases as $\langle T\rangle $ increases. This happens because the FCI is inherently formulated on a lattice and the Goldstone mode acquires a mass because of the presence of explicit breaking of rotational symmetry by the lattice. When $\langle T\rangle $ is positive, but small in absolute value, the mean-field potential is very shallow and quantum fluctuations effectively restore the symmetry, shifting the critical value of $T$ to a positive number and leading to a larger gap. As the value of $\langle T\rangle $ increases, the mean-field potential becomes deeper, the fluctuations become less important, and, ultimately, the gap is set entirely by the lattice effects. It is likely that Ref.~\cite{jackson2015geometric} has numerically observed the nematic phase in FCIs, softened by the lattice effects and the fluctuations around the mean field.

This completes the formulation of the bimetric effective theory. The full action is given by
\be\la{eq:Seff}
S_{\rm eff}[a,\hat g;g] = S_{\rm top}[a,\hat g] + S_{\rm kin}[\hat g;g] + S_{\rm pot}[\hat g;g]\,.
\ee
Note that, by construction, the effective action $S_{\rm eff}[a,\hat g;g]$ is gauge, diffeomorphism, $SO(2)$, and $\widehat{SO}(2)$ invariant. The structure of the action is rigidly fixed by the symmetries and does not allow further freedom in the lowest order in gradients. Table~\ref{table:fields} summarizes the fields that appear in this paper and their transformation laws.

\begin{center}
\begin{table}
\begin{tabular}{cccccccccc}
\hline
\hline
 & ~$e^A_i$~  & ~$g_{ij}$~ & ~$\omega_\mu $~ & ~$\hat e^\alpha_i$~ & ~$\hat g_{ij}$~ & ~$\hat \omega_\mu$~  & ~$\lambda^{\alpha}{}_A$~ & ~$\mathfrak h_{AB}$~ &  ~$C_\mu$  \\
\hline
 $~SO(2)~$ & $+$  & $-$ & $+$  & $-$ & $-$  & $-$ & $+$ & $+$ & $-$ \\ 
 $~\widehat{SO}(2)~$ & ~$-$~  & $-$ & $-$  & $+$ & $-$  & $+$ & $+$ & $-$ & $-$ \\ 
Diff & $+$  & $+$ & $+$  & $+$ & $+$  & $+$ & $-$ & $-$ & $+$ \\ 
$SL(2,\mathbb R)~$ & $-$  & $-$ & $-$  & $+$ & $+$  & $+$ & $+$ & $+$ & $+$ \\ 
\hline\hline
\end{tabular}
\caption{The full list of fields that appear in various formulations of bimetric theory. The ``+'' entries indicate that the field transforms under the symmetry group, while the ``$-$'' entries indicate that it does not.}
\label{table:fields}
\end{table}
\end{center}

\subsection{Hall viscosity}

We start with an attempt to fix the coefficient $\varsigma$ by calculating the Hall viscosity \cite{avron1995viscosity, read2009non} from Eq.~\eqref{eq:Seff}. Hall viscosity describes the response of the stress tensor to time-dependent strain, defined by the generating functional as
\be
T^{\mu}_{\hphantom{\mu}\nu}=\frac{\delta  W[A_\mu, g_{ij}]}{\delta e_\mu^A}e_\nu^A=\lambda^{\mu\hphantom{\nu}\lambda}_{\hphantom{\mu}\nu\hphantom{\lambda}A}e_\lambda^A+\eta^{\mu\hphantom{\nu}\lambda}_{\hphantom{\mu}\nu\hphantom{\lambda}A}\partial_0e_\lambda^A\,.
\ee
The Hall viscosity 
\begin{equation}
{(\eta_H)}^{\mu\hphantom{\nu}\lambda}_{\hphantom{\mu}\nu\hphantom{\lambda}\rho}=
\frac{1}{2}\left(\eta^{\mu\hphantom{\nu}\lambda}_{\hphantom{\mu}\nu\hphantom{\lambda}A}-\eta^{\lambda\hphantom{A}\mu}_{\hphantom{\lambda}A\hphantom{\mu}\nu}\right) e_\rho^A
\end{equation}
is the nondissipative, parity-odd part of the viscosity tensor. When rotational symmetry is preserved, it has one independent component  $\eta_H=\mathcal{S}\bar \rho/4$, proportional to the shift on the sphere \cite{avron1998odd,read2009non},  and the average density $\bar{\rho}$. In the nematic phase, the Hall viscosity will no longer have only one independent component.

To compute the Hall viscosity, we integrate out the gauge field $a$ and find
\begin{multline}
S_{\rm eff}[\hat g;g] = \frac{\nu}{4\pi} \!\int\! AdA + \frac{\nu s}{2\pi} \!\int\! Ad\omega \\ \la{eq:bmunprojected}
+ \frac{\nu \varsigma}{2\pi} \!\int\! Ad\hat \omega + S_{\rm kin}[\hat g;g] + S_{\rm pot}[\hat g;g]\,,
\end{multline}
 where we have dropped the gravitational Chern-Simons terms generated by integration over $a$ \cite{GCYFA}. The first term in the second line of Eq.~\eqref{eq:bmunprojected} was referred to as the ``Berry phase term'' in Refs.\ \cite{maciejko2013field, you2014theory}.

 As discussed previously, at low frequency, the term $S_{\rm pot}[\hat g;g]$ forces the dynamical metric $\hat g_{ij}$ to follow $g_{ij}$, and the constant part $h_{ij}$ is fixed to be $0$ to minimize the potential. Plugging the isotropic solution of the equations of motion $\hat g_{ij} = g_{ij}$, which implies $\hat \omega_i = \omega_i$ \footnote{In the nematic phase, $\omega_i \neq \hat \omega_i$, and the argument no longer holds.}, back into the action, we find
\be
S_{\rm eff}[ g] = \frac{\nu}{4\pi} \!\int\! AdA + \frac{\nu (s + \varsigma)}{2\pi} \!\int\! Ad\omega\,.
\ee
Then, Hall viscosity is given by
\be
\eta_{H} = \frac{2s + 2\varsigma}{4}\bar \rho = \frac{\mathcal S}{4}\bar \rho\,,
\ee<
 where we have introduced $\bar \rho = \frac{\nu}{2\pi} B$. Thus, we were able to match $s+\varsigma$ to the shift, which was already done in Eq.\ \eqref{shift}. We find that the matching of the Hall viscosity is reproduced correctly; however, it cannot be used to fix $\varsigma$.  This is in contrast to Ref.~\cite{maciejko2013field}, where $\varsigma$ was identified with $\mathcal S/2$ by matching the Hall viscosity. It is clear from bimetric theory that the authors of Ref.~\cite{maciejko2013field} implicitly assumed $s=0$. In what follows, we show that in the isotropic phase, $\varsigma$ can indeed be uniquely fixed by matching the projected static structure factor.
 
Note that if the background metric is perturbed at a frequency much larger than the gap and much smaller than the cyclotron frequency $\tilde m\ll \Omega \ll \omega_c$, the dynamic metric $\hat g_{ij}$ can no longer follow $g_{ij}$, and instead we find $\hat g_{ij} = \delta_{ij}$. 
Thus, we conclude that at higher frequencies, the interacting quantum Hall state responds to the metric 
perturbations as if it has shift $\mathcal S = 2s$.
 
 Next, we linearize the theory around a flat background and establish the relation with the previous works \cite{maciejko2013field,you2014theory}.
\subsection{Linearization of bimetric theory}

To gain further insight into the effective theory \eqref{eq:Seff}, we expand the action to the quadratic order in flat space, $g_{ij}= \delta_{ij}$. We assume that the theory is in the isotropic phase $\gamma<1$. 
We choose the following parametrization of $\hat g_{ij}$,
\be
 \hat g_{ij} = \exp
\begin{pmatrix} 
 Q_2 & Q_1\\
Q_1& -Q_2
\end{pmatrix} \approx \begin{pmatrix} 
1+ Q_2 & Q_1\\
Q_1 & 1-Q_2
\end{pmatrix}\,,
\ee
and furthermore,
\be
 Q = Q_1+ iQ_2\,, \qquad  \bar Q = Q_1-iQ_2\,.
\ee
It is not hard to see that $\hat e^\alpha_i=\hat e^\alpha_i(Q)$ depends \emph{nonlinearly} on $Q$ and $\bar Q$; thus, to simplify the analysis, we linearize the theory around the isotropic vacuum, $Q_i=0$.

It is convenient to first integrate out $a$ in Eq.\ \eqref{eq:Seff} to find
\be\la{eq:Seffnoa}
S_{\rm eff}[Q] = \frac{\nu\varsigma}{2\pi} \!\int\! Ad\hat \omega   + S_{\rm kin}[Q] +S_{\rm pot}[Q]\,,
\ee
where we have  removed all of the terms made only from the external fields and neglected the gravitational Chern-Simons term $\hat \omega d \hat \omega$ since it is higher order in derivatives. This term will be discussed in some detail in the next section. We integrate the first term in Eq.\ \eqref{eq:Seffnoa} by parts and expand everything up to the second order in $Q$ to find
\bea
S_{\rm top} &\approx& \frac{i\varsigma \bar \rho}{4} \!\int\!  \bar Q \dot Q ,\\
S_{\rm kin} &\approx& -\alpha \!\int\!  |\partial Q|^2 , \\
S_{\rm pot} &\approx& - \frac{m}{2} \!\int\!|Q|^2\,, \quad m = \tilde m (1-\gamma) ,
\eea
giving us the low-energy, linearized Lagrangian
\be\la{eq:Llinear}
\mathscr L_{\rm eff} \approx \frac{i \varsigma \bar \rho}{4} \bar Q \dot Q - \alpha |\p Q|^2 - \frac{m}{2} |Q|^2\,.  
\ee
The equations of motion for $Q$, 
\be
i\dot Q = -\frac{16\alpha}{\varsigma \bar \rho} \Delta Q + \frac{2m}{\varsigma \bar \rho} Q\,,
\ee
imply that as long as $Q$ tends to $0$ at infinity, the only solution is $Q=0$, which corresponds to $\hat g_{ij} = g_{ij} = \delta_{ij}.$ The action of the type shown in Eq.\ \eqref{eq:Llinear} was studied in Ref.~\cite{maciejko2013field} and derived in Ref.~\cite{you2014theory} using flux attachment. The Lagrangians studied in Refs.~\cite{maciejko2013field,you2014theory} had different values for the coefficient $\varsigma$. We will resolve the disagreement shortly.

The propagator for the field $Q$ is given by
\be
\big \langle \bar Q Q \big \rangle = \frac{4}{\varsigma\bar\rho} \frac{i}{\Omega - \frac{\alpha}{\varsigma \bar\rho}(k\ell)^2 - \frac{2m}{\varsigma \bar \rho }+i0}\,,
\ee
where $\ell = \sqrt{1/B}$ is the magnetic length. 

\subsection{Projected static structure factor}

In this section, we show how to fix the coefficient $\varsigma$. To do so, we evaluate the projected static structure factor, defined as the connected equal-time correlation function of the (projected) density operators,
\be
\bar s (k) =\frac{1}{\bar \rho} \!\int\! \frac{d \Omega}{2\pi}\, \big\langle \rho(\Omega, k) \rho(-\Omega,-k) \big\rangle_c\,.
\ee
We define $s_4$ and $s_6$ as the coefficients in the low-$k$ expansion of $\bar s(k)$,
\begin{equation}\label{eq:s4s6}
  \bar s(k) = s_4 (k\ell)^4 + s_6 (k\ell)^6 + o\left((k\ell)^6\right).
\end{equation}
To evaluate the SSF, we recall Eq.\ \eqref{eq:density},
\be\la{eq:densityflat}
\rho = \bar \rho + \frac{\nu \varsigma}{4\pi} \hat R\,.
\ee
Thus, the SSF is given by the correlation function of Ricci scalars (which are operators in bimetric theory),
\be\la{eq:SSFR}
\bar s(k) = \left(\frac{\nu \varsigma}{4\pi} \right)^2 \frac{1}{\bar \rho} \!\int\! \frac{d \Omega}{2\pi}\, \Big \langle \hat R(\Omega,k) \hat R(-\Omega, -k) \Big \rangle_c\,.
\ee
To evaluate the correlation function, we express the Ricci curvature $\hat R$ in terms of $Q$,
\be
\hat R = 2i (\bar \p^2  Q - \p^2 \bar Q)\,.
\ee
Using the propagator at zero spatial momentum, 
\be
\langle \bar Q Q \rangle = \frac{4}{\bar \rho \varsigma}\, \frac{i}{\Omega - 2\Omega_0+i0}\,,\quad \Omega_0 = \frac m{\varsigma\bar \rho}\,,
\ee
 we find, in momentum space,
\be
\Big \langle \hat R(\Omega,k) \hat R(-\Omega, -k) \Big \rangle_c = 
 \frac{k^4}{\bar\rho\varsigma} \frac{4i\Omega_0}{\Omega^2-(2\Omega_0)^2+i0}
\ee
which leads to
\be
\bar s(k) = \frac{2\varsigma }{8} (k\ell)^4 \,,
\ee
where we find that $2\varsigma = 8s_4$. This is one of the central results of the present paper.

We can match this to a general relation, valid for chiral states in the absence of Landau level mixing \cite{haldane2009hall,nguyen2014lowest} (see also Ref.~\cite{read2011hall}),
\be
\bar s(k) = \frac{\mathcal S-1}{8} (k\ell)^4+\ldots\,,
\ee
from where we fix 
\be
2\varsigma = \mathcal S -1
\ee 
and $2s = 1$. 

The decomposition $\mathcal S = 2\varsigma + 2s$, in the present case, corresponds to the separation of the shift into contributions from the orbital part $s$ and the guiding center part $\varsigma$. This is in contrast to Ref.~\cite{maciejko2013field}, where the coefficient $\varsigma$ was fixed to be equal to $\mathcal S/2$. The realization of this decomposition in the effective theory \eqref{eq:Seff} is one of the main results of the present paper.  The GMP mode is absent in the IQH case (since the coefficient of the $Ad\hat \omega$ term vanishes), which is a crucial property of any effective theory of the GMP mode.

In the language of Ref.~\cite{you2014theory}, the composite fermions (in the isotropic phase) should couple to both the fluctuations of the GMP mode and background geometry through the covariant derivative 
\be
D_i = \p_i + iA_i +ia_i + i\frac{1}{2}\omega_i + i\frac{k-2}{2}\hat \omega_i\,.
\ee
 Thus, the composite fermions perceive the ambient geometry in the same way as particles with a geometric spin $s=\frac{1}{2}$---the same value as for noninteracting electrons filling the lowest Landau level. At the same time, the composite fermions perceive the fluctuations of the GMP mode (viewed in the second-order, or metric, formalism)  as particles with a geometric spin $\varsigma = \frac{k-2}{2} + \frac{1}{2} = \frac{k-1}{2}$, equal to the geometric spin of the guiding centers \cite{haldane2011geometrical}.

 To lowest order in derivatives, such as presented in Eq.~\eqref{eq:Llinear}, there are no more dimensionless adjustable parameters. The phenomenological coefficient $C$ discussed in Ref.~\cite{maciejko2013field} is fixed to the value $-\frac{1}{2}$ by the requirement that the effective theory \eqref{eq:Seff} is self-consistent in a weakly curved space. For any other value of $C$, the theory is not invariant under  the internal local $SO(2)$ ``symmetry'' and would require additional noninvariant counter terms to restore the invariance. These extra terms will enforce $C=-\frac{1}{2}$.

We emphasize that, at this point, we have not proven that $\rho$ should be treated as a projected density operator, although the vanishing of the $(k\ell)^2$ contribution to the SSF is strong evidence. In the next section, we provide further evidence of the implicit LLL projection present in Eq.\ \eqref{eq:Seff}.

\subsection{GMP algebra}

In this section, we prove that the density operators \eqref{eq:densityflat} satisfy the long-wave limit of the GMP/$W_\infty$ algebra \cite{cappelli1993infinite,iso1992fermions}. It was shown in Ref.~\cite{girvin1986magneto} that projected density operators $\rho(\bk)$ satisfy the following commutation relations:
\bea
\nonumber
[\rho(\bk), \rho(\bq)] &=& 2i 
e^{\frac12(\bk \cdot \bq) \ell^2}
\sin\left( \frac{(\bk \times \bq)  \ell^2}{2}\right)\rho(\bk+\bq)\\ \la{eq:GMP}
 &\approx& i (\bk \times \bq) \ell^2 \rho(\bq+\bk)\,.
\eea
 To reproduce the GMP algebra from bimetric theory, we start with the expression \eqref{eq:densityflat}; however, this time we express the linearized spin connection $\hat \omega_i$ and the Ricci scalar $\hat R$ in terms of the field $\hat g_{ij}$:
\be
\hat \omega_i = -\frac{1}{2} \epsilon_{jk} \p_j \hat g_{ki}\,, \qquad \hat R = -\p_i \p_j \hat g_{ij}\,.
\ee
The electron density is given by
\be\la{eq:densityomega}
\rho = \bar \rho + \frac{\nu \varsigma}{2\pi} \epsilon_{ij} \p_i \hat \omega_j\,.
\ee

To evaluate the commutator, we need to know the commutation relations for $\hat g_{ij}$. These commutation relations follow from the commutation relations for $\hat e^\alpha_i$ that can be read out from the topological term of Eq.\ \eqref{eq:Seff}. We find
\be
[\hat E^i_\alpha(\bx), \hat e^\beta_j(\bx^\prime)] = -\frac{2i}{\bar \rho \varsigma}\delta^i_{j}\epsilon_{\alpha}{}^{\beta}\delta(\bx - \bx^\prime)\,.
\ee
Upon linearization in $\hat e^\alpha{}_{i} \approx \delta^{\alpha}_i+ \delta \hat e^\alpha_{i}$, we find 
\be\la{eq:commlambda}
[\delta\hat e^\alpha_i(\bx), \delta\hat e^\beta_j(\bx^\prime)] \approx -\frac{2i}{\bar \rho \varsigma} \delta_{ij} \epsilon^{\alpha\beta} \delta(\bx - \bx^\prime)\,,
\ee
which is identical to the commutation relations obtained in Ref.~\cite{maciejko2013field}. We find that components of  the metric satisfy the $\mathfrak{sl}(2,\mathbb R)$ algebra \cite{haldane2011geometrical, maciejko2013field},
\begin{multline}
[ \hat g_{ij}(\bx), \hat g_{kl}(\bx^\prime)] = -\frac{2i}{\bar \rho \varsigma}\Big( \epsilon_{il} \hat g_{jk} + \epsilon_{jk}\hat g_{il} \\ 
 + \epsilon_{jl}\hat g_{ik} + \epsilon_{ik}\hat g_{jl}\Big)\delta(\bx - \bx^\prime)\,.
\end{multline}

It is useful to evaluate the commutator of spin connections $\hat \omega_i$ as an intermediate step. Curiously, the algebra of $\hat \omega_i$ closes (to lowest order in $k$). We find
\bea\nonumber
[\hat \omega_i(\bk),\, \hat \omega_j(\bq)] &=& \frac{1}{\bar \rho \varsigma}\left(k_j \hat \omega_i(\bk+\bq)  - q_i \hat \omega_j(\bk+\bq)\right)\\ \la{eq:omegacomm}
&- & \frac{i\epsilon_{ij} }{2\bar \rho \varsigma} \hat R(\bk+\bq)\,,
\eea
which implies 
\be\la{eq:riccicomm}
[\hat R(\bk), \hat R(\bq)] =\frac{4\pi}{\nu \varsigma} i (\bk \times \bq) \ell^2\hat R(\bk+\bq)\,.
\ee
Together with $\rho = \bar \rho + \frac{\nu \varsigma}{4\pi} \hat R$, Eq.\ \eqref{eq:riccicomm} implies the GMP algebra \eqref{eq:GMP}. Thus, we are justified in treating the operators $\rho$, at low energies, as the projected density operators. Since the commutators in Eqs.\ \eqref{eq:omegacomm} and \eqref{eq:riccicomm} take a covariant form, we expect that they will remain valid beyond the linear order in $\hat g_{ij}$.  We were not able to find the ``spin connection algebra'' \eqref{eq:omegacomm} in the literature.
Finally, we note here that the GMP algebra has also appeared in fractional Chern insulators \cite{parameswaran2013fractional}, further supporting our claim that bimetric theory can be used to describe the FCIs. 

\section{Further properties of bimetric theory}

In this section, we use bimetric theory to calculate the $s_6$ coefficient of the projected SSF and derive the small-momentum behavior of the GMP mode. 

\subsection{Gravitational Chern-Simons term}

So far, we have neglected the gravitational Chern-Simons term, which will necessarily be generated from integrating out the gauge field. This term has a quantized, dimensionless coefficient and therefore will most likely describe some universal physics. In this section, we fix this coefficient and expose the physical consequences of this term.

 Bimetric theory, modified by the Chern-Simons term, takes the form
\be\la{eq:bm-full}
S_{\rm eff}[\hat g;g] =\frac{\nu\varsigma}{2\pi} \!\int\! Ad\hat \omega + S_{\rm gCS}[\hat g]+ S_{\rm kin}[\hat g;g] + S_{\rm pot}[\hat g;g]  \,,
\ee
where 
\be\la{eq:gcs}
 S_{\rm gCS}[\hat g] = -\frac{\hat c}{4\pi} \int \hat \omega d \hat \omega\,.
\ee
Here, we have introduced a phenomenological parameter $\hat c$ and, for now, will remain agnostic about its value and origin. So far, the only constraint is that $\hat c$ should vanish for the IQH states.

Note that the gravitational Chern-Simons term is third order in derivatives and, therefore, will not modify either the dispersion relation, the projected SSF, or the GMP algebra at leading order. Thus, the role of the term is quite subtle, and to expose it, we go to higher orders in the momentum expansion.

Writing the gravitational Chern-Simons term in components, we find
\be
\int\! \hat \omega d \hat \omega = \int\! \hat \omega_0\hat R - \epsilon_{ij} \hat \omega_i \dot{\hat \omega}_j\,,
\ee
which we linearize around a flat background metric. The first term is inherently nonlinear and goes as $Q^3$; thus, it will be disregarded. The second term, however, contributes to the quadratic effective action
\be
 S_{\rm gCS}[Q] =- \frac{\hat c}{4\pi} \int \hat \omega d \hat \omega \approx -\frac{i\hat c}{16\pi} \int \bar Q \Delta \dot Q\,.
\ee
The gravitational Chern-Simons term $S_{\rm gCS}$ has several subtle consequences.
First, it modifies the canonical commutation relations for $\hat g_{ij}$ by a higher-gradient term \footnote{
The term \protect{$\hat \omega_0 \hat R$} will add a nonlinear contribution to the CCR.
Presently, the effect of this nonlinear correction is not clear to us.}, 
\begin{multline} \la{eq:CCRgCS}
[ \hat g_{ij}(\bk), \hat g_{kl}(\bq)]  =-\frac{2i}{\bar \rho \varsigma}\left(1 - \frac{\hat c}{2\nu \varsigma} \ell^2(\mathbf{k} \cdot \mathbf{q}) \right) \\ \times \Big( \epsilon_{il} \hat g_{jk} + \epsilon_{jk}\hat g_{il} 
 + \epsilon_{jl}\hat g_{ik} + \epsilon_{ik}\hat g_{jl}\Big)\,,
\end{multline}
where every $\hat g_{ij}$ on the rhs is taken at the momentum $\mathbf{k}+\mathbf{q}$. We now verify that the additional term in Eq.~\eqref{eq:CCRgCS} does not spoil the GMP algebra to the subleading, $(k\ell)^4$, order in the momentum expansion. To do so in a self-consistent manner, we add all the terms to bimetric theory that contribute in this order. Fortunately, there is only one such term:
\be
S_{\rm \xi}[\hat g]= \frac{\xi \ell^2}{8\pi} \int d^3x \sqrt{g}\,\, \hat g^{ij} E_i \p_j \hat R\,,
\ee
where $\xi$ is a phenomenological parameter to be fixed momentarily and $E_i$ is the electric field. This term describes the interaction of the dynamic curvature $\hat R$ with a gradient of the electric field, favoring an alignment of the gradient of the dynamical curvature with the electric field.

 The role of this term is to modify the relation between the electron density and Ricci curvature $\hat R$:
\be\la{eq:newdensity}
\rho = \bar \rho + \frac{\nu \varsigma}{4\pi} \hat R + \frac{\xi}{8\pi} \ell^2 \Delta \hat R\,.
\ee
The commutation relation of the density operators is now
\be
[\rho(\bk), \rho(\bq)] = i (\bk \times \bq) \rho -  \frac{i}{8\pi}(\hat c+\xi)  (\bk \times \bq) (\bk \cdot \bq) \hat R\,,
\ee
where $\rho$ is given by Eq.\,\eqref{eq:newdensity}. To ensure that the GMP algebra holds to order $(k\ell)^4$, we enforce 
\be
\xi=-\hat c - \frac{\nu \varsigma}{2}\,.
\ee 
We do not attempt to reproduce the $(k\ell)^6$ term in the GMP algebra since it is likely that, in order to properly match such a term to the exact GMP expression, the higher-spin fields have to be introduced. A recent attempt to find a topological theory for higher-spin fields was made in Ref.~\cite{cappelli2015multipole}.

Next, we fix the coefficient $\hat c$ in terms of microscopic properties of the FQH states. To accomplish this task, we evaluate the $(k\ell)^6$ correction to the projected static structure factor, Eq.\ \eqref{eq:s4s6},
 and compare it to the results known in the literature. Under certain conditions on the microscopic physics, discussed in Refs.\ \cite{nguyen2014lowest, haldane2011geometrical, CLW, can2014geometry, Gromov-galilean, Nguyen-PHD}, $s_6$ is universal and is determined by  topological quantum numbers (i.e., the filling factor, shift, chiral central charge, etc.) only. Generally speaking, we do not expect $s_6$ to be universal because its ``universality'' arises from subtle (and not yet entirely understood) properties of the quantum Hall phase \cite{CLW, Nguyen-PHD}. The necessary (but not sufficient) conditions are (i) the absence of Landau level mixing and that (ii) the state has to be either fully chiral, in the sense of Ref.~\cite{nguyen2014lowest}, or a particle-hole conjugate of a chiral state \cite{Nguyen-PHD}.

When evaluated in bimetric theory, the coefficient $s_6$ receives two different contributions. One comes from the gravitational Chern-Simons term, through the modification of the propagator
\be\la{eq:CSprop}
\big \langle \bar Q Q \big \rangle = \frac{4}{\varsigma\bar\rho} \frac{i}{\left(1 + \frac{\hat c}{2\nu\varsigma} (k\ell)^2\right) \Omega- \frac{\alpha}{2\pi \nu \varsigma}(k\ell)^2  - \frac{2m}{\varsigma \bar \rho} + i0}\,,
\ee
and another one comes from $S_{\rm \xi}[\hat g]$ \footnote{To be really consistent, we have to include higher-spin fields in the theory. These fields will also enter the definition of the density. Instead of doing this, we take a more phenomenological approach; we assume that higher-spin fields have already been integrated out and have generated the correction to the density, proportional to $\Delta \hat R$. }. Computation reveals
\be
\bar s(k) = \frac{2\varsigma}{8} (k\ell)^4 - \frac{\hat c+2\xi}{8\nu} (k\ell)^6 =   \frac{2\varsigma}{8} (k\ell)^4 + \frac{\hat c - \nu \varsigma}{8\nu} (k\ell)^6\,.
\ee
Thus, we establish a general relation $\hat c =  8\nu s_6$.
Assuming the absence of Landau level mixing and chirality of the state, we can match $\hat c$ to Refs.~\cite{CLW, can2014geometry, Nguyen-PHD} and find
\be\la{eq:ch-s6}
\hat c=8\nu s_6 + 4 \nu s_4 =  \frac{\nu-c}{12} + \nu \varsigma^2 + \nu \,{\rm var}(s)\,,
\ee
where $\nu\,{\rm var}(s)$ is the orbital spin variance \cite{GCYFA, bradlyn2015topological}, which plays a major role when the theory is applied to the Jain series. Note that $\hat c$ vanishes for IQH states. At this point, it is good to note that the value of the coefficient $\hat c$ is not the naive one that would have come from integrating out only the gauge field $a$. The extra  contributions should come from integrating out the higher-spin modes. Since the higher-spin modes were not introduced explicitly, we use the consistency of bimetric theory with the GMP algebra and $s_6$ to fix the value of $\hat c$. It is an interesting open problem to perform the computation with the higher-spin fields directly.

The present calculation of $s_6$ can be turned around to argue its universality in the SMA. Indeed, the coefficient $\hat c$ of the gravitational Chern-Simons term cannot take an arbitrary value and has to be quantized as a rational number. At the same time, Eq.\ \eqref{eq:ch-s6} tells us that $s_6 = \frac{\hat c}{8\nu} - \frac{\varsigma}{8}$; thus, if $\hat c$ cannot change continuously under small perturbations (provided that these perturbations do not introduce the LL mixing), then neither can $s_6$. 

A comment is in order. It is well known that any two-body Hamiltonian, projected to the lowest Landau level, is invariant under the particle-hole (PH) transformation. This invariance implies a duality \cite{Nguyen-PHD} between FQH states and their particle-hole conjugates (in the sense of Girvin \cite{girvin1984particle}). 

 We now comment on the property of our theory with respect to PH symmetry \cite{girvin1984particle}, which is an exact symmetry of any two-body Hamiltonian, projected to the lowest Landau level. PH transformation relates the properties of a QH state with its particle-hole conjugate.
If we wish to claim that $S_{\rm eff}[\hat g;g]$ describes the intrinsic dynamics, projected to a single Landau level, then particle-hole duality must be manifest. Upon closer inspection, we find that this is indeed the case. Note that the coefficients $\nu\varsigma$ and $\hat c$ are of the form $\nu s_4$ and $\nu s_6$, respectively. It was shown in Refs.\ \cite{SL,Nguyen-PHD} that the combination $\nu \bar s(k)$ is invariant under PH transformation.
Since the coefficients in the effective action have the form $\nu s_i$ [$s_i$ being the coefficient in front of $(k\ell)^{i}$ in the long-wave expansion of the projected SSF], we conclude that particle-hole duality is manifest in the bimetric effective action $S_{\rm eff}[\hat g;g]$. The actions proposed in Refs.~\cite{maciejko2013field, you2014theory} do not satisfy this property. For completeness, we note that under PH transformation, the first three terms in Eq.~\eqref{eq:bm-full} flip the sign, while the rest of the action is invariant. This transformation property is similar to the one encountered in Dirac composite fermion theory of Jain states, where PH transformation only flips certain signs.

Next, we use Eq.\ \eqref{eq:CSprop} to find the dispersion relation of the GMP mode:
\be\la{eq-GMPGCS}
\frac{\Omega(k)}{\Omega(0)} = 1+ \left(\frac{\alpha}{2m} - \frac{\hat c}{2\nu\varsigma} \right) (k\ell)^2\,.
\ee
The presence of a universal correction in the $(k\ell)^2$ order of the dispersion relation should not come as a surprise. To see that such a possibility exists, we can use the results of Ref.~\cite{girvin1986magneto}, which state that the dispersion relation of the GMP mode, in the single-mode approximation, is given by 
\be
\Omega(k) = \frac{ f(k)}{\bar s (k)}\,,
\ee
where $f(k)$ is a nonuniversal function that depends on the interaction potential \cite{girvin1986magneto}. Assuming that $f(k)$ takes the form  $f(k) = f_4 (k\ell)^4 + f_6 (k\ell)^6 + \ldots$, we find
\be
\frac{\Omega(k)}{\Omega(0)} =  1 + \left( \frac{f_6}{ f_4} - \frac{s_6}{s_4} \right)(k\ell)^2  + \ldots\,.
\ee
Comparing this to Eqs.\ \eqref{eq:ch-s6} and \eqref{eq-GMPGCS}, we unambiguously match the phenomenological parameters
\be\la{eq:wow}
m = 2 \bar \rho f_4\,, \qquad \alpha  = 4 \bar \rho f_6 - 2 \bar \rho f_4\,.
\ee  
Equation~\eqref{eq:wow} provides the explicit expressions of the phenomenological parameters $m$ and $\alpha$ in terms of the Fourier modes of the projected interaction potential, as long as the SMA remains valid. 

The first identification in Eq.\ \eqref{eq:wow} is also consistent with Ref.~\cite{girvin1986magneto}, where the function $f(k)$ is identified with the expectation value of the double commutator $[\rho({\bf k}),[\rho(-{\bf k}),H]]$.  In our theory, $\rho$ is given by the second derivative of the metric $\hat g_{ij}$, while the Hamiltonian is, to leading order, given by Eq.\ \eqref{eq:Spot}. Computing the double commutator, we find the first term in the low-$k$ expansion of $f(k)$, $f_4(k\ell)^4$. This term is sometimes referred to as the ``shear modulus'' \cite{chui1986shear, haldane2011self, golkar2016spectral}.

\subsection{Global properties}
In this section, we briefly discuss some global features of bimetric theory. We start by noting that the coefficient $\hat c$ has two distinct contributions:
\be
\hat c = \frac{\nu-c}{12} +\nu\Big[\varsigma^2+  \,{\rm var}(s)\Big]\,.
\ee
It is now understood \cite{GCYFA} that upon integrating out the gauge field $a$ (and, presumably, the higher-spin fields) in addition to the gravitational Chern-Simons terms, one generates a winding number term $\sim (\hat E d \hat e)^3$. This winding number term is completely invisible as far as any local properties are concerned. It can, however, be observed if one is interested in global properties such as global gravitational anomalies (i.e., noninvariance under Dehn twists). We conjecture that the winding term is indeed generated in bimetric theory and has the form
\be\la{eq:wind}
S_{\rm winding}[\hat e] = \frac{1}{3} \frac{c-\nu}{96\pi} \int \Tr\left( \hat E d \hat e\right)^3\,.
\ee
To support the conjecture, we note that the combination $(c-\nu)/24$ has previously appeared in the numerical computation by Park and Haldane \cite{park2014guiding}, where it appeared as a subleading correction to what they called ``momentum polarization.'' We are not aware of how to make a precise connection, however, we note that when the theory is placed on a (spatial) torus, only the winding number term transforms under the Dehn twist of the cylinder and, therefore, contributes to the phase accumulated by the wave function after the twist, which is the quantity that should be computed in the momentum polarization technique. We note, however, that the Dehn twist must only affect the $\hat g_{ij}$ and not the spatial metric $g_{ij}$. This is, presumably, accomplished in Ref.\,\cite{park2014guiding} via an ``orbital cut.''

 We also emphasize that Eq.\ \eqref{eq:wind} does {\it not} imply that the gravitational anomaly of the edge theory is given by $c-\nu$. Another winding number term, made from the ambient vielbein $e$, is generated from the framing anomaly and ensures the correct gravitational anomaly of the edge.

Finally, we note that taking the limit $m\rightarrow \infty$ leads to the elimination of all local degrees of freedom and pushes the gap of the GMP mode to infinity. In this limit, however, $S_{\rm eff}$ reduces to a topological theory given by $S_{\rm top}$. Singularities of $\hat \omega$ in such a theory correspond to world lines of geometric singularities such as conical points \cite{schine2015synthetic, PhysRevLett.117.266803, klevtsov2016lowest}, punctures \cite{can2016central}, or branching points \cite{gromov2016geometric} of the dynamical metric $\hat g$. The singularities of $\hat \omega$, with the curvature given by
\be
\hat R = 4\pi \frac{p}{2\varsigma}\delta(z-z_0(t)),
\ee
accommodate the proper Wilson lines of the $U(1)_k$ Chern-Simons theory and, consequently, may correspond to quasiholes.

\bigskip
\section{Conclusions}

We have formulated a bimetric theory for the gapped collective excitations in fractional quantum Hall states. The theory consists of topological Chern-Simons theory coupled to a nonrelativistic, parity-violating version of a bimetric massive gravity. The theory is naturally formulated in curved ambient space with gauge, diffeomorphism, and internal $SO(2)$ ``symmetries'' built into the construction. The symmetries rigidly restrict the form of the effective action, leaving only two dimensionless parameters, $\varsigma$ and $\hat c$. There is, in principle, an infinite number of dimensionful parameters, only two of which play a role in the low-energy physics. For the sake of completeness, we write the entire Lagrangian (after the gauge field is ``integrated out'')
\bea\nonumber
\mathscr L_{\rm eff} &=& \frac{\nu\varsigma}{2\pi}  Ad\hat \omega - \frac{\hat c}{4\pi}  \hat \omega d \hat \omega  -\frac{\hat c \ell^2}{8\pi}  \hat g^{ij}E_i \p_j \hat R\\ \nonumber
&-& \frac{\tilde m}{2} \left(\frac{1}{2}\hat g_{ij} g^{ij} - \gamma  \right)^2 - \frac{\alpha}{4}\left| \Gamma - \hat \Gamma \right|^2  \\
&+& \frac{1}{3} \frac{c-\nu}{96\pi}  \Tr\left( \hat E d \hat e\right)^3,
\eea
where $\varsigma = 4s_4$ and $\hat c = 8\nu s_6 + 4\nu s_4$.

 The full bimetric theory is nonlinear. To perform the computations, we linearized the theory around the flat background and found an action similar to the one studied previously \cite{maciejko2013field}; however, we used fewer phenomenological parameters because some of these parameters were fixed by the symmetry and by the consistency of the theory in weakly curved space. 
 The values of the parameters are self-consistent and do not agree with Refs.~\cite{maciejko2013field, you2014theory}, however, they do agree with Refs.~\cite{golkar2016spectral, golkar2016higher} whenever comparison is possible.
 
 We have related the value of $\varsigma$ to the $s_4$ coefficient in the long-wave expansion  of the \emph{projected} static structure factor $\varsigma = 4  s_4$. In the limit when Landau level mixing can be neglected, and the state is chiral (or a particle-hole conjugate of a chiral state) this sets  $2|\varsigma| = |\mathcal S - 1|$, where $\mathcal S$ is the shift. This value of $\varsigma$ implies the absence of the GMP mode in the integer case (as it should be).

 The dimensionless coefficient $\hat c$ appears in front of the gravitational Chern-Simons term and determines the $s_6$ coefficient in the long-wave expansion of the projected static structure factor according to $\hat c = - 8\nu s_6$; it contributes to the $(k\ell)^2$ behavior of the GMP dispersion relation at low wavelength. 
 
 The projected static structure factor is interpreted in bimetric theory as a two-point function of Ricci scalars. The GMP algebra follows from the fact that components of the spin connection $\hat \omega_i$ form a closed algebra themselves. In principle, any FQH calculation performed in bimetric theory receives a curious geometric interpretation.
 
 The present theory is only a first step in the effective field theory description of the bulk gapped collective excitations in quantum Hall states, and many open questions remain. For example, the theory \eqref{eq:Seff} is nonlinear, and it would be interesting to understand the role of the nonlinear effects. Recently, it was understood that many modes of higher angular momentum are expected to appear in addition to the GMP mode. These modes should be described by the higher-spin cousins of the field $\hat g_{ij}$. Presumably, when all of the higher-spin fields are included, the exact GMP algebra should be reproduced. Multilayer states can provide a natural FQH interpretation of multigravity (where many metrics are involved).  Our construction was only spatially covariant, and a fully covariant formulation should be done in the language of the Newton-Cartan geometry. The Galilean or Milne boost symmetry (in the limit of zero bare electron mass $m_{el} \to 0$) is most likely present in bimetric theory, but it is not clear how it is realized. A precise relation to the theory of fractional Chern insulators still needs to be established. Are there higher-spin fields relevant for the FCIs? What is the precise relation to the nonlinear collective field theory of Ref.~\cite{laskin2015collective}? We leave all of these and many other questions to future work.
 
 \acknowledgments It is our pleasure to thank A.\ Abanov, B.\ Bradlyn, P.\ Glorioso, D.\ Haldane, D.\ X.\ Nguyen, Z.\  Papic, R.\ Roy, J.\ Wang, and B.\ Yang for stimulating discussions, and A.\ Abanov, S.\ Moroz, and S.\ Klevtsov for comments on the manuscript. A.\ G.\ also thanks B.\ Bradlyn and S.\ Geraedts for a collaboration on a closely related topic \cite{AGBB}. This work is supported, in part, by U.S.\ DOE Grant
No.\ DE-FG02-13ER41958, the ARO MURI Grant No.\ 63834-PH-MUR, and a
Simons Investigator Grant from the Simons Foundation.  Additional
support was provided by the Chicago MRSEC, which is funded by NSF
through Grant No.\ DMR-1420709. A.\ G. was supported by
NSF Grant No.\ DMS-1206648.
 
\appendix
\section{Values of the coefficients for specific states}
In this appendix, we collect the values of the topological quantum numbers for some prominent states. We have found that the phenomenological coefficients are given by the coefficients of small momentum expansion of the projected static structure factor $\bar s(q) = s_4 q^4 + s_4 q^6+ \ldots$ according to $\varsigma = 4s_4$ and $\hat c = 8 \nu s_6$, which implies 
\bea
\varsigma &=& \frac{\mathcal S-1}{2} \, ,\\
\hat c &=&  \frac{\nu-c}{12} - \nu \varsigma + \nu \varsigma^2 + \nu \,{\rm var}(s) \, .
\eea
The values of the coefficients for various states are summarized in the table below.
\begin{widetext}
\begin{center}
\begin{tabular}{ |c|c|c|c|c|c|c| } 
\hline
& & & & & & \\[-1em]
 & $\nu$ & $\mathcal S$ & $c$ & $\nu{\rm var}(s) $& $\varsigma$ & $\hat c$ \\[.5em]
\hline
& & & & & & \\[-.75em]
Laughlin\,\cite{laughlin1983anomalous} & $\displaystyle{\frac{1}{k}}$ & $k$ & $1$ & $0$ & $~\displaystyle{\frac{k-1}{2}}~$ &  $\displaystyle{\frac{(k-1)(3 k-10)}{12 k}}$\\[1em]
\hline
& & & & & & \\[-.75em]
$~$Moore-Read\,\cite{moore1991nonabelions}$~$ & $\displaystyle{\frac{1}{2}}$ & $3$ & $~\displaystyle{\frac{3}{2}}~$ & $0$ & $1$ & $-\displaystyle{\frac{1}{12}}$\\[1em] 
\hline
& & & & & & \\[-.75em]
Read-Rezayi\,\cite{Read1999} & $~\displaystyle{\frac{k}{Mk+2}}~$ & $~M+2~$ & $\displaystyle{\frac{3k}{k+2}}$ & $0$ & $\displaystyle{\frac{M+1}{2}}$ & $~~\displaystyle{\frac{k \left(3 (k+2) M^2-3 k M-2 (k+5)\right)}{12 (k+2) (k M+2)}}~~$  \\[1em]
\hline
& & & & & & \\[-.75em]
Jain\,\cite{jain1989composite} & $\displaystyle{\frac{N}{2N+1}}$ & $N+2$ & $N$ & $~\displaystyle{\frac{N(N^2-1)}{12}}~$ & $~\displaystyle{\frac{N+1}{2}}~$ & $\displaystyle{\frac{N \left(N^3+2 N^2-2 N-2\right)}{6 (2 N+1)}}$ \\[1em]
\hline
& & & & & & \\[-.75em]
Jain, $p$ fluxes& $\displaystyle{\frac{N}{2Np+1}}$ & $N+2p$ & $N$ & $~\displaystyle{\frac{N(N^2-1)}{12}}~$ & $~\displaystyle{\frac{N+2p-1}{2}}~$ & $\displaystyle{\frac{N \left(N^3 p +2 N^2+N(4p{-}6)+6p(p{-}2) +4\right)}{6(2 N p+1)}}$ \\[1em]
\hline
\end{tabular}
\end{center}
\end{widetext}

The topological quantum numbers for the particle-hole dual states can be found in Ref.~\cite{Nguyen-PHD}. The coefficients $\varsigma$ and $\hat c$ do not change under  particle-hole transformation. Note that the coefficient $\hat c<0$ only for the Read-Rezayi series at $M=1$ and, in particular, for the Laughlin $\nu=\frac{1}{3}$ and $\nu = \frac{2}{3}$ states and for the Pfaffian and anti-Pfaffian $\nu=\frac{1}{2}$ states. The value of $\hat c$ for the Pfaffian state leads to a gravitational Chern-Simons term with a properly quantized coefficient $\frac{1}{48\pi} \hat \omega d \hat \omega $.
\section{Potential term in bimetric gravity}
\label{app:potential}

The most prominent feature of bimetric gravity is the fine-tuned potential that ensures the absence of ghosts. Here, we construct this potential in the language of bimetric theory. For simplicity, we assume (as it is usually done in bimetric gravity) that there is no difference between $SO(2)$ and $\widehat{SO}(2)$ indices. To construct the potential, we follow Ref.~\cite{bimetric2010} and introduce
\be
\gamma^i{}_j[g,\hat g] = g^{ik} \hat g_{kj}= e^A_j\mathfrak h_A{}^B E^i_B\,, \quad \mathcal K^i{}_j[g,\hat g] = \delta^i_j - (\sqrt{\gamma})^i{}_j\,.
\ee
It then follows that
\be
\mathcal K^i{}_j[g,\hat g] =  E^i_A  \Big( \delta^A_B - \lambda^A{}_B\Big) e_j^B\,.
\ee
Given these definitions, there is a unique stable potential \cite{bimetric2010}
\begin{multline}\la{eq:Spotapp}
S_{\rm pot} = m \!\int\! d^3x\, \sqrt{g} \Big( [\mathcal K]^2-  [\mathcal K^2] \Big) \\
= 2m \!\int\! d^3x\, \sqrt{g}\Big( 1 + \lambda - [\lambda] \Big)  \,,
\end{multline}
where we have introduced $[\mathcal K] = \Tr \mathcal K$, $\lambda = \det \lambda$, and $m$ is a phenomenological parameter that will determine the gap of the GMP mode. This potential supports only \emph{one} phase. As long as $m>0$, there is a unique vacuum state with unbroken rotational symmetry. Indeed, in the parametrization $Q_1 = \rho \cos\phi$, $Q_2 = \rho \sin \phi$. We find
\be
S_{\rm pot} = -8m \!\int\! d^3 x\, (\sinh \rho)^2\,,
\ee
which means that, as long as $\rho$ is real, the minimum occurs at $\rho = 0$ which implies $h_{ij} = 0$ and $\lambda^A{}_B=\delta^A_B$ for all positive $m$. So, there is no possibility of a phase transition.

\section{Linearized geometric quantities}
\label{app:linearized}
In this appendix, we collect several useful formulas for linearized geometric objects. We parametrize the $\mh_{AB}$ as 
\be
\mh_{AB}  \approx \begin{pmatrix} 
1+ Q_2 & Q_1\\
Q_1 & 1-Q_2
\end{pmatrix} .
\ee
We use the following notations:
\be
Q= Q_1 + iQ_2\,, \qquad \bar Q  = Q_1 - i Q_2,
\ee
and
\be
\p = \frac{1}{2} (\p_x - i \p_y)\,, \qquad \bar \p = \frac{1}{2} (\p_x +i\p_y)\,,
\ee
so that $\p z =\bar \p \bar z =1$ and $\bar \p z = \p \bar z =0$. Under internal $SO(2)$ rotation by angle $\varphi$, the fields transform according to $Q\rightarrow e^{-2i\varphi}Q$ and $\bar Q\rightarrow e^{2i\varphi}\bar Q$. The angle is \emph{twice} the angle of rotation of an arbitrary vector $v_z$ because $Q$ has a quadrupolar nature.

The spin connection is given by
\be
\hat \omega_{i} = -\frac{1}{2} \epsilon_{jk} \p_j \hat g_{ki}\,.
\ee
We define the complex spin connection as
\be
\hat \omega_z = \omega_1 + i\omega_2\,, \qquad \hat \omega_{\bar z} = \omega_{1} - i\omega_2\,.
\ee
Then
\be
\hat \omega_z =  \p  \bar Q \,,\qquad \hat \omega_{\bar z} =   \bar \p  Q\,.
\ee
Next, we compute $C_k = \epsilon^i{}_j\hat\Gamma^j{}_{i,k} = 2\hat \omega_k$.

Thus, it is clear that <
\be
\mathscr L_{\rm kin} = - \frac{\alpha}{4}  C_i  C_i = - \frac{\alpha}{4}  C_z  C_{\bar z} = - \alpha |\p Q|^2\,.
\ee
We find the Ricci curvature
\be
\hat R = - \p_i \p_j \hat g_{ij} = 2i (\bar \p^2  Q - \p^2 \bar Q)\,.
\ee
We also find the ``gravi-electric'' field
\be
\hat{\mathcal E}_z = \p_0 \hat \omega_z\,, \qquad \hat{\mathcal E}_{\bar z} = \p_0 \hat \omega_{\bar z}\,.
\ee
The temporal component of the spin connection is always quadratic in fields,
\be
\hat \omega_0 = \frac{1}{2} \epsilon^i{}_j \hat E{}^j \p_0 \hat e^k{}_i \approx  \frac{i}{8}\left( Q \dot{\bar Q}-\bar Q \dot Q\right)\,.
\ee
The temporal component of $C_0$ is also quadratic in fields; however, it is written entirely in terms of $\hat G^{ij}$ and $\hat g_{ij}$:
\be
C_0 = \frac{1}{2} \epsilon_{i}{}^j \hat G^{ik}\p_0\hat g_{jk}\,.
\ee
While, upon linearization, this term is indistinguishable from $\hat \omega_0$, we remark that as far as the canonical commutation relations are concerned, this term modifies the commutator of $[\hat g_{ij},\hat g_{kl}]$ by a term that is zeroth order in $\hat g_{ij}$. Then, $\hat g_{ij}$ do not form an $sl(2,\mathbb R)$ algebra. For this reason, we forbid the coupling of the Chern-Simons theory to $C_\mu$ and do not include the term $\bar \rho C_0$ in the effective action $S_{\rm eff}$.

The gravitational Chern-Simons integrand is given by
\be
\hat \omega d \hat \omega \approx -\epsilon_{ij} \hat \omega_i \p_0 \hat \omega_j \approx \frac{i}{2}\left( \bar \p Q \p \dot{\bar Q} - \p\bar Q   \bar \p \dot Q \right) .
\ee
We recall that $\p \bar \p = \frac{1}{4} \Delta$, where $\Delta$ is the Laplace operator. The Chern-Simons term, as emphasized in the main text, modifies the canonical commutation relations.

\bibliography{Bibliography-proof}

\begin{thebibliography}{75}%
\makeatletter
\providecommand \@ifxundefined [1]{%
 \@ifx{#1\undefined}
}%
\providecommand \@ifnum [1]{%
 \ifnum #1\expandafter \@firstoftwo
 \else \expandafter \@secondoftwo
 \fi
}%
\providecommand \@ifx [1]{%
 \ifx #1\expandafter \@firstoftwo
 \else \expandafter \@secondoftwo
 \fi
}%
\providecommand \natexlab [1]{#1}%
\providecommand \enquote  [1]{``#1''}%
\providecommand \bibnamefont  [1]{#1}%
\providecommand \bibfnamefont [1]{#1}%
\providecommand \citenamefont [1]{#1}%
\providecommand \href@noop [0]{\@secondoftwo}%
\providecommand \href [0]{\begingroup \@sanitize@url \@href}%
\providecommand \@href[1]{\@@startlink{#1}\@@href}%
\providecommand \@@href[1]{\endgroup#1\@@endlink}%
\providecommand \@sanitize@url [0]{\catcode `\\12\catcode `\$12\catcode
  `\&12\catcode `\#12\catcode `\^12\catcode `\_12\catcode `\%12\relax}%
\providecommand \@@startlink[1]{}%
\providecommand \@@endlink[0]{}%
\providecommand \url  [0]{\begingroup\@sanitize@url \@url }%
\providecommand \@url [1]{\endgroup\@href {#1}{\urlprefix }}%
\providecommand \urlprefix  [0]{URL }%
\providecommand \Eprint [0]{\href }%
\providecommand \doibase [0]{http://dx.doi.org/}%
\providecommand \selectlanguage [0]{\@gobble}%
\providecommand \bibinfo  [0]{\@secondoftwo}%
\providecommand \bibfield  [0]{\@secondoftwo}%
\providecommand \translation [1]{[#1]}%
\providecommand \BibitemOpen [0]{}%
\providecommand \bibitemStop [0]{}%
\providecommand \bibitemNoStop [0]{.\EOS\space}%
\providecommand \EOS [0]{\spacefactor3000\relax}%
\providecommand \BibitemShut  [1]{\csname bibitem#1\endcsname}%
\let\auto@bib@innerbib\@empty
\bibitem [{\citenamefont {Tsui}\ \emph {et~al.}(1982)\citenamefont {Tsui},
  \citenamefont {Stormer},\ and\ \citenamefont {Gossard}}]{Tsui:1982yy}%
  \BibitemOpen
  \bibfield  {author} {\bibinfo {author} {\bibfnamefont {D.~C.}\ \bibnamefont
  {Tsui}}, \bibinfo {author} {\bibfnamefont {H.~L.}\ \bibnamefont {Stormer}}, \
  and\ \bibinfo {author} {\bibfnamefont {A.~C.}\ \bibnamefont {Gossard}},\
  }\bibfield  {title} {\enquote {\bibinfo {title} {{Two-Dimensional
  Magnetotransport in the Extreme Quantum Limit}},}\ }\href {\doibase
  10.1103/PhysRevLett.48.1559} {\bibfield  {journal} {\bibinfo  {journal}
  {Phys. Rev. Lett.}\ }\textbf {\bibinfo {volume} {48}},\ \bibinfo {pages}
  {1559} (\bibinfo {year} {1982})}\BibitemShut {NoStop}%
\bibitem [{\citenamefont {Laughlin}(1983)}]{laughlin1983anomalous}%
  \BibitemOpen
  \bibfield  {author} {\bibinfo {author} {\bibfnamefont {R.~B.}\ \bibnamefont
  {Laughlin}},\ }\bibfield  {title} {\enquote {\bibinfo {title} {{Anomalous
  Quantum Hall Effect: An Incompressible Quantum Fluid with Fractionally
  Charged Excitations}},}\ }\href {\doibase 10.1103/PhysRevLett.50.1395}
  {\bibfield  {journal} {\bibinfo  {journal} {Phys. Rev. Lett.}\ }\textbf
  {\bibinfo {volume} {50}},\ \bibinfo {pages} {1395} (\bibinfo {year}
  {1983})}\BibitemShut {NoStop}%
\bibitem [{\citenamefont {Zhang}\ \emph {et~al.}(1989)\citenamefont {Zhang},
  \citenamefont {Hansson},\ and\ \citenamefont {Kivelson}}]{ZHK1989}%
  \BibitemOpen
  \bibfield  {author} {\bibinfo {author} {\bibfnamefont {S.~C.}\ \bibnamefont
  {Zhang}}, \bibinfo {author} {\bibfnamefont {T.~H.}\ \bibnamefont {Hansson}},
  \ and\ \bibinfo {author} {\bibfnamefont {S.}~\bibnamefont {Kivelson}},\
  }\bibfield  {title} {\enquote {\bibinfo {title} {{Effective-Field-Theory
  Model for the Fractional Quantum Hall Effect}},}\ }\href {\doibase
  10.1103/PhysRevLett.62.82} {\bibfield  {journal} {\bibinfo  {journal} {Phys.
  Rev. Lett.}\ }\textbf {\bibinfo {volume} {62}},\ \bibinfo {pages} {82}
  (\bibinfo {year} {1989})}\BibitemShut {NoStop}%
\bibitem [{\citenamefont {Read}(1989)}]{read1989order}%
  \BibitemOpen
  \bibfield  {author} {\bibinfo {author} {\bibfnamefont {N.}~\bibnamefont
  {Read}},\ }\bibfield  {title} {\enquote {\bibinfo {title} {{Order Parameter
  and Ginzburg-Landau Theory for the Fractional Quantum Hall Effect}},}\ }\href
  {\doibase 10.1103/PhysRevLett.62.86} {\bibfield  {journal} {\bibinfo
  {journal} {Phys. Rev. Lett.}\ }\textbf {\bibinfo {volume} {62}},\ \bibinfo
  {pages} {86} (\bibinfo {year} {1989})}\BibitemShut {NoStop}%
\bibitem [{\citenamefont {Moore}\ and\ \citenamefont
  {Read}(1991)}]{moore1991nonabelions}%
  \BibitemOpen
  \bibfield  {author} {\bibinfo {author} {\bibfnamefont {G.}~\bibnamefont
  {Moore}}\ and\ \bibinfo {author} {\bibfnamefont {N.}~\bibnamefont {Read}},\
  }\bibfield  {title} {\enquote {\bibinfo {title} {{Nonabelions in the
  Fractional Quantum Hall Effect}},}\ }\href {\doibase
  10.1016/0550-3213(91)90407-O} {\bibfield  {journal} {\bibinfo  {journal}
  {Nucl. Phys.}\ }\textbf {\bibinfo {volume} {B360}},\ \bibinfo {pages} {362}
  (\bibinfo {year} {1991})}\BibitemShut {NoStop}%
\bibitem [{\citenamefont {Haldane}(1983)}]{haldane1983fractional}%
  \BibitemOpen
  \bibfield  {author} {\bibinfo {author} {\bibfnamefont {F.~D.~M.}\
  \bibnamefont {Haldane}},\ }\bibfield  {title} {\enquote {\bibinfo {title}
  {{Fractional Quantization of the Hall Effect: A Hierarchy of Incompressible
  Quantum Fluid States}},}\ }\href {\doibase 10.1103/PhysRevLett.51.605}
  {\bibfield  {journal} {\bibinfo  {journal} {Phys. Rev. Lett.}\ }\textbf
  {\bibinfo {volume} {51}},\ \bibinfo {pages} {605} (\bibinfo {year}
  {1983})}\BibitemShut {NoStop}%
\bibitem [{\citenamefont {Wen}\ and\ \citenamefont
  {Zee}(1992)}]{WenZeeShiftPaper}%
  \BibitemOpen
  \bibfield  {author} {\bibinfo {author} {\bibfnamefont {X.~G.}\ \bibnamefont
  {Wen}}\ and\ \bibinfo {author} {\bibfnamefont {A.}~\bibnamefont {Zee}},\
  }\bibfield  {title} {\enquote {\bibinfo {title} {{Shift and Spin Vector: New
  Topological Quantum Numbers for the Hall Fluids}},}\ }\href {\doibase
  10.1103/PhysRevLett.69.953} {\bibfield  {journal} {\bibinfo  {journal} {Phys.
  Rev. Lett.}\ }\textbf {\bibinfo {volume} {69}},\ \bibinfo {pages} {953}
  (\bibinfo {year} {1992})}\BibitemShut {NoStop}%
\bibitem [{\citenamefont {Fr\"ohlich}\ and\ \citenamefont
  {Studer}(1993)}]{1993-frohlich}%
  \BibitemOpen
  \bibfield  {author} {\bibinfo {author} {\bibfnamefont {J.}~\bibnamefont
  {Fr\"ohlich}}\ and\ \bibinfo {author} {\bibfnamefont {U.~M.}\ \bibnamefont
  {Studer}},\ }\bibfield  {title} {\enquote {\bibinfo {title} {{Gauge
  Invariance and Current Algebra in Nonrelativistic Many-Body Theory}},}\
  }\href {\doibase 10.1103/RevModPhys.65.733} {\bibfield  {journal} {\bibinfo
  {journal} {Rev. Mod. Phys.}\ }\textbf {\bibinfo {volume} {65}},\ \bibinfo
  {pages} {733} (\bibinfo {year} {1993})}\BibitemShut {NoStop}%
\bibitem [{\citenamefont {Avron}\ \emph {et~al.}(1995)\citenamefont {Avron},
  \citenamefont {Seiler},\ and\ \citenamefont {Zograf}}]{avron1995viscosity}%
  \BibitemOpen
  \bibfield  {author} {\bibinfo {author} {\bibfnamefont {J.~E.}\ \bibnamefont
  {Avron}}, \bibinfo {author} {\bibfnamefont {R.}~\bibnamefont {Seiler}}, \
  and\ \bibinfo {author} {\bibfnamefont {P.~G.}\ \bibnamefont {Zograf}},\
  }\bibfield  {title} {\enquote {\bibinfo {title} {{Viscosity of Quantum Hall
  Fluids}},}\ }\href {\doibase 10.1103/PhysRevLett.75.697} {\bibfield
  {journal} {\bibinfo  {journal} {Phys. Rev. Lett.}\ }\textbf {\bibinfo
  {volume} {75}},\ \bibinfo {pages} {697} (\bibinfo {year} {1995})}\BibitemShut
  {NoStop}%
\bibitem [{\citenamefont {Tokatly}\ and\ \citenamefont
  {Vignale}(2007)}]{tokatly2007lorentz}%
  \BibitemOpen
  \bibfield  {author} {\bibinfo {author} {\bibfnamefont {I.~V.}\ \bibnamefont
  {Tokatly}}\ and\ \bibinfo {author} {\bibfnamefont {G.}~\bibnamefont
  {Vignale}},\ }\bibfield  {title} {\enquote {\bibinfo {title} {{Lorentz Shear
  Modulus of a Two-Dimensional Electron Gas at High Magnetic Field}},}\ }\href
  {\doibase 10.1103/PhysRevB.76.161305} {\bibfield  {journal} {\bibinfo
  {journal} {Phys. Rev. B}\ }\textbf {\bibinfo {volume} {76}},\ \bibinfo
  {pages} {161305} (\bibinfo {year} {2007})}\BibitemShut {NoStop}%
\bibitem [{\citenamefont {Read}(2009)}]{read2009non}%
  \BibitemOpen
  \bibfield  {author} {\bibinfo {author} {\bibfnamefont {N.}~\bibnamefont
  {Read}},\ }\bibfield  {title} {\enquote {\bibinfo {title} {{Non-Abelian
  Adiabatic Statistics and Hall Viscosity in Quantum Hall States and $p_x+ i
  p_y$ Paired Superfluids}},}\ }\href {\doibase 10.1103/PhysRevB.79.045308}
  {\bibfield  {journal} {\bibinfo  {journal} {Phys. Rev. B}\ }\textbf {\bibinfo
  {volume} {79}},\ \bibinfo {pages} {045308} (\bibinfo {year}
  {2009})}\BibitemShut {NoStop}%
\bibitem [{\citenamefont {Haldane}()}]{haldane2009hall}%
  \BibitemOpen
  \bibfield  {author} {\bibinfo {author} {\bibfnamefont {F.~D.~M.}\
  \bibnamefont {Haldane}},\ }\bibfield  {title} {\enquote {\bibinfo {title}
  {{``Hall viscosity'' and Intrinsic Metric of Incompressible Fractional Hall
  Fluids}},}\ }\href@noop {} {\ }\Eprint {http://arxiv.org/abs/0906.1854}
  {arXiv:0906.1854} \BibitemShut {NoStop}%
\bibitem [{\citenamefont {Abanov}\ and\ \citenamefont
  {Gromov}(2014)}]{Abanov-2014}%
  \BibitemOpen
  \bibfield  {author} {\bibinfo {author} {\bibfnamefont {A.~G.}\ \bibnamefont
  {Abanov}}\ and\ \bibinfo {author} {\bibfnamefont {A.}~\bibnamefont
  {Gromov}},\ }\bibfield  {title} {\enquote {\bibinfo {title} {{Electromagnetic
  and Gravitational Responses of Two-Dimensional Noninteracting Electrons in a
  Background Magnetic Field}},}\ }\href {\doibase 10.1103/PhysRevB.90.014435}
  {\bibfield  {journal} {\bibinfo  {journal} {Phys. Rev. B}\ }\textbf {\bibinfo
  {volume} {90}},\ \bibinfo {pages} {014435} (\bibinfo {year}
  {2014})}\BibitemShut {NoStop}%
\bibitem [{\citenamefont {Klevtsov}(2014)}]{klevtsov2014random}%
  \BibitemOpen
  \bibfield  {author} {\bibinfo {author} {\bibfnamefont {S.}~\bibnamefont
  {Klevtsov}},\ }\bibfield  {title} {\enquote {\bibinfo {title} {{Random Normal
  Matrices, Bergman Kernel and Projective Embeddings}},}\ }\href {\doibase
  10.1007/JHEP01(2014)133} {\bibfield  {journal} {\bibinfo  {journal} {J. High
  Energy Phys.}\ }\textbf {\bibinfo {volume} {01}},\ \bibinfo {pages} {133}
  (\bibinfo {year} {2014})}\BibitemShut {NoStop}%
\bibitem [{\citenamefont {Ferrari}\ and\ \citenamefont
  {Klevtsov}(2014)}]{ferrari2014fqhe}%
  \BibitemOpen
  \bibfield  {author} {\bibinfo {author} {\bibfnamefont {F.}~\bibnamefont
  {Ferrari}}\ and\ \bibinfo {author} {\bibfnamefont {S.}~\bibnamefont
  {Klevtsov}},\ }\bibfield  {title} {\enquote {\bibinfo {title} {{FQHE on
  Curved Backgrounds, Free Fields and Large N}},}\ }\href {\doibase
  10.1007/JHEP12(2014)086} {\bibfield  {journal} {\bibinfo  {journal} {J. of
  High Energy Phys.}\ }\textbf {\bibinfo {volume} {12}},\ \bibinfo {pages}
  {086} (\bibinfo {year} {2014})}\BibitemShut {NoStop}%
\bibitem [{\citenamefont {Can}\ \emph {et~al.}(2014)\citenamefont {Can},
  \citenamefont {Laskin},\ and\ \citenamefont {Wiegmann}}]{CLW}%
  \BibitemOpen
  \bibfield  {author} {\bibinfo {author} {\bibfnamefont {T.}~\bibnamefont
  {Can}}, \bibinfo {author} {\bibfnamefont {M.}~\bibnamefont {Laskin}}, \ and\
  \bibinfo {author} {\bibfnamefont {P.}~\bibnamefont {Wiegmann}},\ }\bibfield
  {title} {\enquote {\bibinfo {title} {{Fractional Quantum Hall Effect in a
  Curved Space: Gravitational Anomaly and Electromagnetic Response}},}\ }\href
  {\doibase 10.1103/PhysRevLett.113.046803} {\bibfield  {journal} {\bibinfo
  {journal} {Phys. Rev. Lett.}\ }\textbf {\bibinfo {volume} {113}},\ \bibinfo
  {pages} {046803} (\bibinfo {year} {2014})}\BibitemShut {NoStop}%
\bibitem [{\citenamefont {Gromov}\ \emph {et~al.}(2015)\citenamefont {Gromov},
  \citenamefont {Cho}, \citenamefont {You}, \citenamefont {Abanov},\ and\
  \citenamefont {Fradkin}}]{GCYFA}%
  \BibitemOpen
  \bibfield  {author} {\bibinfo {author} {\bibfnamefont {A.}~\bibnamefont
  {Gromov}}, \bibinfo {author} {\bibfnamefont {G.~Young}\ \bibnamefont {Cho}},
  \bibinfo {author} {\bibfnamefont {Y.}~\bibnamefont {You}}, \bibinfo {author}
  {\bibfnamefont {A.~G.}\ \bibnamefont {Abanov}}, \ and\ \bibinfo {author}
  {\bibfnamefont {E.}~\bibnamefont {Fradkin}},\ }\bibfield  {title} {\enquote
  {\bibinfo {title} {{Framing Anomaly in the Effective Theory of the Fractional
  Quantum Hall Effect}},}\ }\href {\doibase 10.1103/PhysRevLett.114.016805}
  {\bibfield  {journal} {\bibinfo  {journal} {Phys. Rev. Lett.}\ }\textbf
  {\bibinfo {volume} {114}},\ \bibinfo {pages} {016805} (\bibinfo {year}
  {2015})}\BibitemShut {NoStop}%
\bibitem [{\citenamefont {Bradlyn}\ and\ \citenamefont
  {Read}(2015{\natexlab{a}})}]{bradlyn2015topological}%
  \BibitemOpen
  \bibfield  {author} {\bibinfo {author} {\bibfnamefont {B.}~\bibnamefont
  {Bradlyn}}\ and\ \bibinfo {author} {\bibfnamefont {N.}~\bibnamefont {Read}},\
  }\bibfield  {title} {\enquote {\bibinfo {title} {{Topological Central Charge
  from Berry Curvature: Gravitational Anomalies in Trial Wave Functions for
  Topological Phases}},}\ }\href {\doibase 10.1103/PhysRevB.91.165306}
  {\bibfield  {journal} {\bibinfo  {journal} {Phys. Rev. B}\ }\textbf {\bibinfo
  {volume} {91}},\ \bibinfo {pages} {165306} (\bibinfo {year}
  {2015}{\natexlab{a}})}\BibitemShut {NoStop}%
\bibitem [{\citenamefont {Klevtsov}\ and\ \citenamefont
  {Wiegmann}(2015)}]{klevtsov2015precise}%
  \BibitemOpen
  \bibfield  {author} {\bibinfo {author} {\bibfnamefont {S.}~\bibnamefont
  {Klevtsov}}\ and\ \bibinfo {author} {\bibfnamefont {P.}~\bibnamefont
  {Wiegmann}},\ }\bibfield  {title} {\enquote {\bibinfo {title} {{Geometric
  Adiabatic Transport in Quantum Hall States}},}\ }\href {\doibase
  10.1103/PhysRevLett.115.086801} {\bibfield  {journal} {\bibinfo  {journal}
  {Phys. Rev. Lett.}\ }\textbf {\bibinfo {volume} {115}},\ \bibinfo {pages}
  {086801} (\bibinfo {year} {2015})}\BibitemShut {NoStop}%
\bibitem [{\citenamefont {Girvin}\ \emph {et~al.}(1986)\citenamefont {Girvin},
  \citenamefont {MacDonald},\ and\ \citenamefont
  {Platzman}}]{girvin1986magneto}%
  \BibitemOpen
  \bibfield  {author} {\bibinfo {author} {\bibfnamefont {S.~M.}\ \bibnamefont
  {Girvin}}, \bibinfo {author} {\bibfnamefont {A.~H.}\ \bibnamefont
  {MacDonald}}, \ and\ \bibinfo {author} {\bibfnamefont {P.~M.}\ \bibnamefont
  {Platzman}},\ }\bibfield  {title} {\enquote {\bibinfo {title} {{Magneto-roton
  Theory of Collective Excitations in the Fractional Quantum Hall Effect}},}\
  }\href {\doibase 10.1103/PhysRevB.33.2481} {\bibfield  {journal} {\bibinfo
  {journal} {Phys. Rev. B}\ }\textbf {\bibinfo {volume} {33}},\ \bibinfo
  {pages} {2481} (\bibinfo {year} {1986})}\BibitemShut {NoStop}%
\bibitem [{\citenamefont {Hirjibehedin}\ \emph {et~al.}(2005)\citenamefont
  {Hirjibehedin}, \citenamefont {Dujovne}, \citenamefont {Pinczuk},
  \citenamefont {Dennis}, \citenamefont {Pfeiffer},\ and\ \citenamefont
  {West}}]{Pinczuk}%
  \BibitemOpen
  \bibfield  {author} {\bibinfo {author} {\bibfnamefont {C.~F.}\ \bibnamefont
  {Hirjibehedin}}, \bibinfo {author} {\bibfnamefont {I.}~\bibnamefont
  {Dujovne}}, \bibinfo {author} {\bibfnamefont {A.}~\bibnamefont {Pinczuk}},
  \bibinfo {author} {\bibfnamefont {B.~S.}\ \bibnamefont {Dennis}}, \bibinfo
  {author} {\bibfnamefont {L.~N.}\ \bibnamefont {Pfeiffer}}, \ and\ \bibinfo
  {author} {\bibfnamefont {K.~W.}\ \bibnamefont {West}},\ }\bibfield  {title}
  {\enquote {\bibinfo {title} {{Splitting of Long-Wavelength Modes of the
  Fractional Quantum Hall Liquid at $\ensuremath{\nu}=1/3$}},}\ }\href
  {\doibase 10.1103/PhysRevLett.95.066803} {\bibfield  {journal} {\bibinfo
  {journal} {Phys. Rev. Lett.}\ }\textbf {\bibinfo {volume} {95}},\ \bibinfo
  {pages} {066803} (\bibinfo {year} {2005})}\BibitemShut {NoStop}%
\bibitem [{\citenamefont {Kang}\ \emph {et~al.}(2001)\citenamefont {Kang},
  \citenamefont {Pinczuk}, \citenamefont {Dennis}, \citenamefont {Pfeiffer},\
  and\ \citenamefont {West}}]{Pinczuk2}%
  \BibitemOpen
  \bibfield  {author} {\bibinfo {author} {\bibfnamefont {M.}~\bibnamefont
  {Kang}}, \bibinfo {author} {\bibfnamefont {A.}~\bibnamefont {Pinczuk}},
  \bibinfo {author} {\bibfnamefont {B.~S.}\ \bibnamefont {Dennis}}, \bibinfo
  {author} {\bibfnamefont {L.~N.}\ \bibnamefont {Pfeiffer}}, \ and\ \bibinfo
  {author} {\bibfnamefont {K.~W.}\ \bibnamefont {West}},\ }\bibfield  {title}
  {\enquote {\bibinfo {title} {{Observation of Multiple Magnetorotons in the
  Fractional Quantum Hall Effect}},}\ }\href {\doibase
  10.1103/PhysRevLett.86.2637} {\bibfield  {journal} {\bibinfo  {journal}
  {Phys. Rev. Lett.}\ }\textbf {\bibinfo {volume} {86}},\ \bibinfo {pages}
  {2637} (\bibinfo {year} {2001})}\BibitemShut {NoStop}%
\bibitem [{\citenamefont {Kukushkin}\ \emph {et~al.}(2009)\citenamefont
  {Kukushkin}, \citenamefont {Smet}, \citenamefont {Scarola}, \citenamefont
  {Umansky},\ and\ \citenamefont {von Klitzing}}]{kukushkin2009dispersion}%
  \BibitemOpen
  \bibfield  {author} {\bibinfo {author} {\bibfnamefont {I.~V.}\ \bibnamefont
  {Kukushkin}}, \bibinfo {author} {\bibfnamefont {J.~H.}\ \bibnamefont {Smet}},
  \bibinfo {author} {\bibfnamefont {V.~W.}\ \bibnamefont {Scarola}}, \bibinfo
  {author} {\bibfnamefont {V.}~\bibnamefont {Umansky}}, \ and\ \bibinfo
  {author} {\bibfnamefont {K.}~\bibnamefont {von Klitzing}},\ }\bibfield
  {title} {\enquote {\bibinfo {title} {{Dispersion of the Excitations of
  Fractional Quantum Hall States}},}\ }\href {\doibase 10.1126/science.1171472}
  {\bibfield  {journal} {\bibinfo  {journal} {Science}\ }\textbf {\bibinfo
  {volume} {324}},\ \bibinfo {pages} {1044} (\bibinfo {year}
  {2009})}\BibitemShut {NoStop}%
\bibitem [{\citenamefont
  {Haldane}(2011{\natexlab{a}})}]{haldane2011geometrical}%
  \BibitemOpen
  \bibfield  {author} {\bibinfo {author} {\bibfnamefont {F.~D.~M.}\
  \bibnamefont {Haldane}},\ }\bibfield  {title} {\enquote {\bibinfo {title}
  {{Geometrical Description of the Fractional Quantum Hall Effect}},}\ }\href
  {\doibase 10.1103/PhysRevLett.107.116801} {\bibfield  {journal} {\bibinfo
  {journal} {Phys. Rev. Lett.}\ }\textbf {\bibinfo {volume} {107}},\ \bibinfo
  {pages} {116801} (\bibinfo {year} {2011}{\natexlab{a}})}\BibitemShut
  {NoStop}%
\bibitem [{\citenamefont {Haldane}(2011{\natexlab{b}})}]{haldane2011self}%
  \BibitemOpen
  \bibfield  {author} {\bibinfo {author} {\bibfnamefont {F.~D.~M.}\
  \bibnamefont {Haldane}},\ }\bibfield  {title} {\enquote {\bibinfo {title}
  {{Self-Duality and Long-Wavelength Behavior of the Landau-Level
  Guiding-Center Structure Function, and the Shear Modulus of Fractional
  Quantum Hall Fluids}},}\ }\href@noop {} {\  (\bibinfo {year}
  {2011}{\natexlab{b}})},\ \Eprint {http://arxiv.org/abs/1112.0990}
  {arXiv:1112.0990} \BibitemShut {NoStop}%
\bibitem [{\citenamefont {Maciejko}\ \emph {et~al.}(2013)\citenamefont
  {Maciejko}, \citenamefont {Hsu}, \citenamefont {Kivelson}, \citenamefont
  {Park},\ and\ \citenamefont {Sondhi}}]{maciejko2013field}%
  \BibitemOpen
  \bibfield  {author} {\bibinfo {author} {\bibfnamefont {J.}~\bibnamefont
  {Maciejko}}, \bibinfo {author} {\bibfnamefont {B.}~\bibnamefont {Hsu}},
  \bibinfo {author} {\bibfnamefont {S.~A.}\ \bibnamefont {Kivelson}}, \bibinfo
  {author} {\bibfnamefont {Y.}~\bibnamefont {Park}}, \ and\ \bibinfo {author}
  {\bibfnamefont {S.~L.}\ \bibnamefont {Sondhi}},\ }\bibfield  {title}
  {\enquote {\bibinfo {title} {{Field Theory of the Quantum Hall Nematic
  Transition}},}\ }\href {\doibase 10.1103/PhysRevB.88.125137} {\bibfield
  {journal} {\bibinfo  {journal} {Phys. Rev. B}\ }\textbf {\bibinfo {volume}
  {88}},\ \bibinfo {pages} {125137} (\bibinfo {year} {2013})}\BibitemShut
  {NoStop}%
\bibitem [{\citenamefont {You}\ \emph {et~al.}(2014)\citenamefont {You},
  \citenamefont {Cho},\ and\ \citenamefont {Fradkin}}]{you2014theory}%
  \BibitemOpen
  \bibfield  {author} {\bibinfo {author} {\bibfnamefont {Y.}~\bibnamefont
  {You}}, \bibinfo {author} {\bibfnamefont {G.~Y.}\ \bibnamefont {Cho}}, \ and\
  \bibinfo {author} {\bibfnamefont {E.}~\bibnamefont {Fradkin}},\ }\bibfield
  {title} {\enquote {\bibinfo {title} {{Theory of Nematic Fractional Quantum
  Hall States}},}\ }\href {\doibase 10.1103/PhysRevX.4.041050} {\bibfield
  {journal} {\bibinfo  {journal} {Phys. Rev. X}\ }\textbf {\bibinfo {volume}
  {4}},\ \bibinfo {pages} {041050} (\bibinfo {year} {2014})}\BibitemShut
  {NoStop}%
\bibitem [{\citenamefont {Regnault}\ \emph {et~al.}()\citenamefont {Regnault},
  \citenamefont {Maciejko}, \citenamefont {Kivelson},\ and\ \citenamefont
  {Sondhi}}]{regnault2016evidence}%
  \BibitemOpen
  \bibfield  {author} {\bibinfo {author} {\bibfnamefont {N}~\bibnamefont
  {Regnault}}, \bibinfo {author} {\bibfnamefont {J}~\bibnamefont {Maciejko}},
  \bibinfo {author} {\bibfnamefont {S.~A.}\ \bibnamefont {Kivelson}}, \ and\
  \bibinfo {author} {\bibfnamefont {S.~L.}\ \bibnamefont {Sondhi}},\ }\bibfield
   {title} {\enquote {\bibinfo {title} {{Evidence of a Fractional Quantum Hall
  Nematic Phase in a Microscopic Model}},}\ }\href@noop {} {\ }\Eprint
  {http://arxiv.org/abs/1607.02178} {arXiv:1607.02178} \BibitemShut {NoStop}%
\bibitem [{\citenamefont {Xia}\ \emph {et~al.}(2011)\citenamefont {Xia},
  \citenamefont {Eisenstein}, \citenamefont {Pfeiffer},\ and\ \citenamefont
  {West}}]{xia2011evidence}%
  \BibitemOpen
  \bibfield  {author} {\bibinfo {author} {\bibfnamefont {J.}~\bibnamefont
  {Xia}}, \bibinfo {author} {\bibfnamefont {J.~P.}\ \bibnamefont {Eisenstein}},
  \bibinfo {author} {\bibfnamefont {L.~N.}\ \bibnamefont {Pfeiffer}}, \ and\
  \bibinfo {author} {\bibfnamefont {K.~W.}\ \bibnamefont {West}},\ }\bibfield
  {title} {\enquote {\bibinfo {title} {{Evidence for a Fractionally Quantized
  Hall State with Anisotropic Longitudinal Transport}},}\ }\href {\doibase
  10.1038/nphys2118} {\bibfield  {journal} {\bibinfo  {journal} {Nat. Phys.}\
  }\textbf {\bibinfo {volume} {7}},\ \bibinfo {pages} {845} (\bibinfo {year}
  {2011})}\BibitemShut {NoStop}%
\bibitem [{\citenamefont {Golkar}\ \emph
  {et~al.}(2016{\natexlab{a}})\citenamefont {Golkar}, \citenamefont {Nguyen},\
  and\ \citenamefont {Son}}]{golkar2016spectral}%
  \BibitemOpen
  \bibfield  {author} {\bibinfo {author} {\bibfnamefont {S.}~\bibnamefont
  {Golkar}}, \bibinfo {author} {\bibfnamefont {D.~X.}\ \bibnamefont {Nguyen}},
  \ and\ \bibinfo {author} {\bibfnamefont {D.~T.}\ \bibnamefont {Son}},\
  }\bibfield  {title} {\enquote {\bibinfo {title} {{Spectral Sum Rules and
  Magneto-roton as Emergent Graviton in Fractional Quantum Hall Effect}},}\
  }\href {\doibase 10.1007/JHEP01(2016)021} {\bibfield  {journal} {\bibinfo
  {journal} {J. High Energy Phys.}\ }\textbf {\bibinfo {volume} {01}},\
  \bibinfo {pages} {021} (\bibinfo {year} {2016}{\natexlab{a}})}\BibitemShut
  {NoStop}%
\bibitem [{\citenamefont {Yang}(2013)}]{yang2013geometry}%
  \BibitemOpen
  \bibfield  {author} {\bibinfo {author} {\bibfnamefont {K.}~\bibnamefont
  {Yang}},\ }\bibfield  {title} {\enquote {\bibinfo {title} {{Geometry of
  Compressible and Incompressible Quantum Hall States: Application to
  Anisotropic Composite-Fermion Liquids}},}\ }\href {\doibase
  10.1103/PhysRevB.88.241105} {\bibfield  {journal} {\bibinfo  {journal} {Phys.
  Rev. B}\ }\textbf {\bibinfo {volume} {88}},\ \bibinfo {pages} {241105(R)}
  (\bibinfo {year} {2013})}\BibitemShut {NoStop}%
\bibitem [{\citenamefont {Yang}(2016)}]{yang2016acoustic}%
  \BibitemOpen
  \bibfield  {author} {\bibinfo {author} {\bibfnamefont {K.}~\bibnamefont
  {Yang}},\ }\bibfield  {title} {\enquote {\bibinfo {title} {{Acoustic Wave
  Absorption as a Probe of Dynamical Geometrical Response of Fractional Quantum
  Hall Liquids}},}\ }\href {\doibase 10.1103/PhysRevB.93.161302} {\bibfield
  {journal} {\bibinfo  {journal} {Phys. Rev. B}\ }\textbf {\bibinfo {volume}
  {93}},\ \bibinfo {pages} {161302(R)} (\bibinfo {year} {2016})}\BibitemShut
  {NoStop}%
\bibitem [{\citenamefont {Gromov}\ \emph {et~al.}()\citenamefont {Gromov},
  \citenamefont {Geraedts},\ and\ \citenamefont {Bradlyn}}]{AGBB}%
  \BibitemOpen
  \bibfield  {author} {\bibinfo {author} {\bibfnamefont {A.}~\bibnamefont
  {Gromov}}, \bibinfo {author} {\bibfnamefont {S.~D.}\ \bibnamefont
  {Geraedts}}, \ and\ \bibinfo {author} {\bibfnamefont {B.}~\bibnamefont
  {Bradlyn}},\ }\bibfield  {title} {\enquote {\bibinfo {title} {{Investigating
  Anisotropic Quantum Hall States with Bi-metric Geometry}},}\ }\href@noop {}
  {\ }\Eprint {http://arxiv.org/abs/1703.01304} {arXiv:1703.01304} \BibitemShut
  {NoStop}%
\bibitem [{\citenamefont {de~Rham}\ \emph {et~al.}(2011)\citenamefont
  {de~Rham}, \citenamefont {Gabadadze},\ and\ \citenamefont
  {Tolley}}]{bimetric2010}%
  \BibitemOpen
  \bibfield  {author} {\bibinfo {author} {\bibfnamefont {C.}~\bibnamefont
  {de~Rham}}, \bibinfo {author} {\bibfnamefont {G.}~\bibnamefont {Gabadadze}},
  \ and\ \bibinfo {author} {\bibfnamefont {A.~J.}\ \bibnamefont {Tolley}},\
  }\bibfield  {title} {\enquote {\bibinfo {title} {{Resummation of Massive
  Gravity}},}\ }\href {\doibase 10.1103/PhysRevLett.106.231101} {\bibfield
  {journal} {\bibinfo  {journal} {Phys. Rev. Lett.}\ }\textbf {\bibinfo
  {volume} {106}},\ \bibinfo {pages} {231101} (\bibinfo {year}
  {2011})}\BibitemShut {NoStop}%
\bibitem [{\citenamefont {Bergshoeff}\ \emph {et~al.}()\citenamefont
  {Bergshoeff}, \citenamefont {de~Haan}, \citenamefont {Hohm}, \citenamefont
  {Merbis},\ and\ \citenamefont {Townsend}}]{bergshoeff2013zwei}%
  \BibitemOpen
  \bibfield  {author} {\bibinfo {author} {\bibfnamefont {E.~A.}\ \bibnamefont
  {Bergshoeff}}, \bibinfo {author} {\bibfnamefont {S.}~\bibnamefont {de~Haan}},
  \bibinfo {author} {\bibfnamefont {O.}~\bibnamefont {Hohm}}, \bibinfo {author}
  {\bibfnamefont {W.}~\bibnamefont {Merbis}}, \ and\ \bibinfo {author}
  {\bibfnamefont {P.~K}\ \bibnamefont {Townsend}},\ }\bibfield  {title}
  {\enquote {\bibinfo {title} {{Zwei-Dreibein Gravity: A Two-Frame-Field Model
  of 3D Massive Gravity}},}\ }\href {\doibase 10.1103/PhysRevLett.111.111102}
  {\bibinfo  {journal} {Phys. Rev. Lett.}\ ,\ \bibinfo {pages}
  {111102}}\BibitemShut {NoStop}%
\bibitem [{\citenamefont {Platzman}\ and\ \citenamefont
  {He}(1996)}]{platzman1996resonant}%
  \BibitemOpen
\bibfield  {journal} {  }\bibfield  {author} {\bibinfo {author} {\bibfnamefont
  {P.~M.}\ \bibnamefont {Platzman}}\ and\ \bibinfo {author} {\bibfnamefont
  {S.}~\bibnamefont {He}},\ }\bibfield  {title} {\enquote {\bibinfo {title}
  {{Resonant Raman Scattering from Magneto Rotons in the Fractional Quantum
  Hall Liquid}},}\ }\href {\doibase 10.1088/0031-8949/1996/T66/030} {\bibfield
  {journal} {\bibinfo  {journal} {Phys. Scr.}\ }\textbf {\bibinfo {volume}
  {T96}},\ \bibinfo {pages} {167} (\bibinfo {year} {1996})}\BibitemShut
  {NoStop}%
\bibitem [{\citenamefont {Repellin}\ \emph {et~al.}(2014)\citenamefont
  {Repellin}, \citenamefont {Neupert}, \citenamefont
  {Papi\ifmmode~\acute{c}\else \'{c}\fi{}},\ and\ \citenamefont
  {Regnault}}]{Papic-SMA}%
  \BibitemOpen
  \bibfield  {author} {\bibinfo {author} {\bibfnamefont {C.}~\bibnamefont
  {Repellin}}, \bibinfo {author} {\bibfnamefont {T.}~\bibnamefont {Neupert}},
  \bibinfo {author} {\bibfnamefont {Z.}~\bibnamefont
  {Papi\ifmmode~\acute{c}\else \'{c}\fi{}}}, \ and\ \bibinfo {author}
  {\bibfnamefont {N.}~\bibnamefont {Regnault}},\ }\bibfield  {title} {\enquote
  {\bibinfo {title} {{Single-Mode Approximation for Fractional Chern Insulators
  and the Fractional Quantum Hall Effect on the Torus}},}\ }\href {\doibase
  10.1103/PhysRevB.90.045114} {\bibfield  {journal} {\bibinfo  {journal} {Phys.
  Rev. B}\ }\textbf {\bibinfo {volume} {90}},\ \bibinfo {pages} {045114}
  (\bibinfo {year} {2014})}\BibitemShut {NoStop}%
\bibitem [{\citenamefont {Jolicoeur}(2017)}]{jolicoeur2017shape}%
  \BibitemOpen
  \bibfield  {author} {\bibinfo {author} {\bibfnamefont {T.}~\bibnamefont
  {Jolicoeur}},\ }\bibfield  {title} {\enquote {\bibinfo {title} {{Shape of the
  Magnetoroton at $\nu= 1/3$ and $\nu= 7/3$ in Real Samples}},}\ }\href
  {\doibase 10.1103/PhysRevB.95.075201} {\bibfield  {journal} {\bibinfo
  {journal} {Phys. Rev. B}\ }\textbf {\bibinfo {volume} {95}},\ \bibinfo
  {pages} {075201} (\bibinfo {year} {2017})}\BibitemShut {NoStop}%
\bibitem [{\citenamefont {Hoyos}\ and\ \citenamefont
  {Son}(2012)}]{2012-HoyosSon}%
  \BibitemOpen
  \bibfield  {author} {\bibinfo {author} {\bibfnamefont {C.}~\bibnamefont
  {Hoyos}}\ and\ \bibinfo {author} {\bibfnamefont {D.~T.}\ \bibnamefont
  {Son}},\ }\bibfield  {title} {\enquote {\bibinfo {title} {{Hall Viscosity and
  Electromagnetic Response}},}\ }\href {\doibase
  10.1103/PhysRevLett.108.066805} {\bibfield  {journal} {\bibinfo  {journal}
  {Phys. Rev. Lett.}\ }\textbf {\bibinfo {volume} {108}},\ \bibinfo {pages}
  {066805} (\bibinfo {year} {2012})}\BibitemShut {NoStop}%
\bibitem [{\citenamefont {Son}()}]{son2013newton}%
  \BibitemOpen
  \bibfield  {author} {\bibinfo {author} {\bibfnamefont {D.~T.}\ \bibnamefont
  {Son}},\ }\bibfield  {title} {\enquote {\bibinfo {title} {{Newton-Cartan
  Geometry and the Quantum Hall Effect}},}\ }\href@noop {} {\ }\Eprint
  {http://arxiv.org/abs/1306.0638} {arXiv:1306.0638} \BibitemShut {NoStop}%
\bibitem [{\citenamefont {Bradlyn}\ and\ \citenamefont
  {Read}(2015{\natexlab{b}})}]{bradlyn2014low}%
  \BibitemOpen
  \bibfield  {author} {\bibinfo {author} {\bibfnamefont {B.}~\bibnamefont
  {Bradlyn}}\ and\ \bibinfo {author} {\bibfnamefont {N.}~\bibnamefont {Read}},\
  }\bibfield  {title} {\enquote {\bibinfo {title} {{Low-Energy Effective Theory
  in the Bulk for Transport in a Topological Phase}},}\ }\href {\doibase
  10.1103/PhysRevB.91.125303} {\bibfield  {journal} {\bibinfo  {journal} {Phys.
  Rev. B}\ }\textbf {\bibinfo {volume} {91}},\ \bibinfo {pages} {125303}
  (\bibinfo {year} {2015}{\natexlab{b}})}\BibitemShut {NoStop}%
\bibitem [{\citenamefont {Gromov}\ and\ \citenamefont
  {Abanov}(2015)}]{gromov-thermal}%
  \BibitemOpen
  \bibfield  {author} {\bibinfo {author} {\bibfnamefont {A.}~\bibnamefont
  {Gromov}}\ and\ \bibinfo {author} {\bibfnamefont {A.~G.}\ \bibnamefont
  {Abanov}},\ }\bibfield  {title} {\enquote {\bibinfo {title} {{Thermal Hall
  Effect and Geometry with Torsion}},}\ }\href {\doibase
  10.1103/PhysRevLett.114.016802} {\bibfield  {journal} {\bibinfo  {journal}
  {Phys. Rev. Lett.}\ }\textbf {\bibinfo {volume} {114}},\ \bibinfo {pages}
  {016802} (\bibinfo {year} {2015})}\BibitemShut {NoStop}%
\bibitem [{\citenamefont {Jensen}()}]{Jensen2014coupling}%
  \BibitemOpen
  \bibfield  {author} {\bibinfo {author} {\bibfnamefont {K.}~\bibnamefont
  {Jensen}},\ }\bibfield  {title} {\enquote {\bibinfo {title} {{On the Coupling
  of Galilean-Invariant Field Theories to Curved Spacetime}},}\ }\href@noop {}
  {\ }\Eprint {http://arxiv.org/abs/1408.6855} {arXiv:1408.6855} \BibitemShut
  {NoStop}%
\bibitem [{\citenamefont {Gromov}\ \emph {et~al.}(2016)\citenamefont {Gromov},
  \citenamefont {Jensen},\ and\ \citenamefont {Abanov}}]{gromov2016boundary}%
  \BibitemOpen
  \bibfield  {author} {\bibinfo {author} {\bibfnamefont {A.}~\bibnamefont
  {Gromov}}, \bibinfo {author} {\bibfnamefont {K.}~\bibnamefont {Jensen}}, \
  and\ \bibinfo {author} {\bibfnamefont {A.~G.}\ \bibnamefont {Abanov}},\
  }\bibfield  {title} {\enquote {\bibinfo {title} {{Boundary Effective Action
  for Quantum Hall States}},}\ }\href {\doibase 10.1103/PhysRevLett.116.126802}
  {\bibfield  {journal} {\bibinfo  {journal} {Phys. Rev. Lett.}\ }\textbf
  {\bibinfo {volume} {116}},\ \bibinfo {pages} {126802} (\bibinfo {year}
  {2016})}\BibitemShut {NoStop}%
\bibitem [{\citenamefont {de~Rham}(2014)}]{de2014massive}%
  \BibitemOpen
  \bibfield  {author} {\bibinfo {author} {\bibfnamefont {C.}~\bibnamefont
  {de~Rham}},\ }\bibfield  {title} {\enquote {\bibinfo {title} {{Massive
  Gravity}},}\ }\href {\doibase 10.12942/lrr-2014-7} {\bibfield  {journal}
  {\bibinfo  {journal} {Living Rev. Relativity}\ }\textbf {\bibinfo {volume}
  {17}},\ \bibinfo {pages} {7} (\bibinfo {year} {2014})}\BibitemShut {NoStop}%
\bibitem [{Note1()}]{Note1}%
  \BibitemOpen
  \bibinfo {note} {We have explicitly assumed that Chern-Simons fields do not
  couple to the one-form $C$. Such coupling may be included and is discussed in
  Appendix \ref {app:linearized}.}\BibitemShut {Stop}%
\bibitem [{Note2()}]{Note2}%
  \BibitemOpen
  \bibinfo {note} {In curved space, the presence of constant $h_{ij}$ leads to
  a breakdown of diffeomorphism invariance}\BibitemShut {NoStop}%
\bibitem [{\citenamefont {Roy}(2014)}]{roy2014band}%
  \BibitemOpen
  \bibfield  {author} {\bibinfo {author} {\bibfnamefont {R.}~\bibnamefont
  {Roy}},\ }\bibfield  {title} {\enquote {\bibinfo {title} {{Band Geometry of
  Fractional Topological Insulators}},}\ }\href {\doibase
  10.1103/PhysRevB.90.165139} {\bibfield  {journal} {\bibinfo  {journal} {Phys.
  Rev. B}\ }\textbf {\bibinfo {volume} {90}},\ \bibinfo {pages} {165139}
  (\bibinfo {year} {2014})}\BibitemShut {NoStop}%
\bibitem [{\citenamefont {Jackson}\ \emph {et~al.}(2015)\citenamefont
  {Jackson}, \citenamefont {M{\"o}ller},\ and\ \citenamefont
  {Roy}}]{jackson2015geometric}%
  \BibitemOpen
  \bibfield  {author} {\bibinfo {author} {\bibfnamefont {T.~S.}\ \bibnamefont
  {Jackson}}, \bibinfo {author} {\bibfnamefont {G.}~\bibnamefont {M{\"o}ller}},
  \ and\ \bibinfo {author} {\bibfnamefont {R.}~\bibnamefont {Roy}},\ }\bibfield
   {title} {\enquote {\bibinfo {title} {{Geometric Stability of Topological
  Lattice Phases}},}\ }\href {\doibase 10.1038/ncomms9629} {\bibfield
  {journal} {\bibinfo  {journal} {Nat. Commun.}\ }\textbf {\bibinfo {volume}
  {6}},\ \bibinfo {pages} {8629} (\bibinfo {year} {2015})}\BibitemShut
  {NoStop}%
\bibitem [{\citenamefont {Avron}(1998)}]{avron1998odd}%
  \BibitemOpen
  \bibfield  {author} {\bibinfo {author} {\bibfnamefont {J.~E.}\ \bibnamefont
  {Avron}},\ }\bibfield  {title} {\enquote {\bibinfo {title} {{Odd
  Viscosity}},}\ }\href {\doibase 10.1023/A:1023084404080} {\bibfield
  {journal} {\bibinfo  {journal} {J. Stat. Phys.}\ }\textbf {\bibinfo {volume}
  {92}},\ \bibinfo {pages} {543} (\bibinfo {year} {1998})}\BibitemShut
  {NoStop}%
\bibitem [{Note3()}]{Note3}%
  \BibitemOpen
  \bibinfo {note} {In the nematic phase, $\omega _i \not =\protect \mathaccentV
  {hat}05E\omega _i$, and the argument no longer holds.}\BibitemShut {Stop}%
\bibitem [{\citenamefont {Nguyen}\ \emph {et~al.}()\citenamefont {Nguyen},
  \citenamefont {Son},\ and\ \citenamefont {Wu}}]{nguyen2014lowest}%
  \BibitemOpen
  \bibfield  {author} {\bibinfo {author} {\bibfnamefont {D.~X.}\ \bibnamefont
  {Nguyen}}, \bibinfo {author} {\bibfnamefont {D.~T.}\ \bibnamefont {Son}}, \
  and\ \bibinfo {author} {\bibfnamefont {C.}~\bibnamefont {Wu}},\ }\bibfield
  {title} {\enquote {\bibinfo {title} {{Lowest Landau Level Stress Tensor and
  Structure Factor of Trial Quantum Hall Wave Functions}},}\ }\href@noop {} {\
  }\Eprint {http://arxiv.org/abs/1411.3316} {arXiv:1411.3316} \BibitemShut
  {NoStop}%
\bibitem [{\citenamefont {Read}\ and\ \citenamefont
  {Rezayi}(2011)}]{read2011hall}%
  \BibitemOpen
  \bibfield  {author} {\bibinfo {author} {\bibfnamefont {N.}~\bibnamefont
  {Read}}\ and\ \bibinfo {author} {\bibfnamefont {E.~H.}\ \bibnamefont
  {Rezayi}},\ }\bibfield  {title} {\enquote {\bibinfo {title} {{Hall Viscosity,
  Orbital Spin, and Geometry: Paired Superfluids and Quantum Hall Systems}},}\
  }\href {\doibase 10.1103/PhysRevB.84.085316} {\bibfield  {journal} {\bibinfo
  {journal} {Phys. Rev. B}\ }\textbf {\bibinfo {volume} {84}},\ \bibinfo
  {pages} {085316} (\bibinfo {year} {2011})}\BibitemShut {NoStop}%
\bibitem [{\citenamefont {Cappelli}\ \emph {et~al.}(1993)\citenamefont
  {Cappelli}, \citenamefont {Trugenberger},\ and\ \citenamefont
  {Zemba}}]{cappelli1993infinite}%
  \BibitemOpen
  \bibfield  {author} {\bibinfo {author} {\bibfnamefont {A.}~\bibnamefont
  {Cappelli}}, \bibinfo {author} {\bibfnamefont {C.~A.}\ \bibnamefont
  {Trugenberger}}, \ and\ \bibinfo {author} {\bibfnamefont {G.~R.}\
  \bibnamefont {Zemba}},\ }\bibfield  {title} {\enquote {\bibinfo {title}
  {{Infinite Symmetry in the Quantum Hall Effect}},}\ }\href {\doibase
  10.1016/0550-3213(93)90660-H} {\bibfield  {journal} {\bibinfo  {journal}
  {Nucl. Phys.}\ }\textbf {\bibinfo {volume} {B396}},\ \bibinfo {pages} {465}
  (\bibinfo {year} {1993})}\BibitemShut {NoStop}%
\bibitem [{\citenamefont {Iso}\ \emph {et~al.}(1992)\citenamefont {Iso},
  \citenamefont {Karabali},\ and\ \citenamefont {Sakita}}]{iso1992fermions}%
  \BibitemOpen
  \bibfield  {author} {\bibinfo {author} {\bibfnamefont {S.}~\bibnamefont
  {Iso}}, \bibinfo {author} {\bibfnamefont {D.}~\bibnamefont {Karabali}}, \
  and\ \bibinfo {author} {\bibfnamefont {B.}~\bibnamefont {Sakita}},\
  }\bibfield  {title} {\enquote {\bibinfo {title} {{Fermions in the Lowest
  Landau Level. Bosonization, $W_\infty$ algebra, Droplets, Chiral Bosons}},}\
  }\href {\doibase 10.1016/0370-2693(92)90816-M} {\bibfield  {journal}
  {\bibinfo  {journal} {Phys. Lett. B}\ }\textbf {\bibinfo {volume} {296}},\
  \bibinfo {pages} {143} (\bibinfo {year} {1992})}\BibitemShut {NoStop}%
\bibitem [{\citenamefont {Parameswaran}\ \emph {et~al.}(2013)\citenamefont
  {Parameswaran}, \citenamefont {Roy},\ and\ \citenamefont
  {Sondhi}}]{parameswaran2013fractional}%
  \BibitemOpen
  \bibfield  {author} {\bibinfo {author} {\bibfnamefont {S.~A.}\ \bibnamefont
  {Parameswaran}}, \bibinfo {author} {\bibfnamefont {R.}~\bibnamefont {Roy}}, \
  and\ \bibinfo {author} {\bibfnamefont {S.~L.}\ \bibnamefont {Sondhi}},\
  }\bibfield  {title} {\enquote {\bibinfo {title} {{Fractional Quantum Hall
  Physics in Topological Flat Bands}},}\ }\href {\doibase
  10.1016/j.crhy.2013.04.003} {\bibfield  {journal} {\bibinfo  {journal} {C. R.
  Phys.}\ }\textbf {\bibinfo {volume} {14}},\ \bibinfo {pages} {816} (\bibinfo
  {year} {2013})}\BibitemShut {NoStop}%
\bibitem [{Note4()}]{Note4}%
  \BibitemOpen
  \bibinfo {note} {The term \protect {$\protect \mathaccentV {hat}05E\omega _0
  \protect \mathaccentV {hat}05ER$} will add a nonlinear contribution to the
  CCR. Presently, the effect of this nonlinear correction is not clear to
  us.}\BibitemShut {Stop}%
\bibitem [{\citenamefont {Cappelli}\ and\ \citenamefont
  {Randellini}(2016)}]{cappelli2015multipole}%
  \BibitemOpen
  \bibfield  {author} {\bibinfo {author} {\bibfnamefont {A.}~\bibnamefont
  {Cappelli}}\ and\ \bibinfo {author} {\bibfnamefont {E.}~\bibnamefont
  {Randellini}},\ }\bibfield  {title} {\enquote {\bibinfo {title} {{Multipole
  Expansion in the Quantum Hall Effect}},}\ }\href {\doibase
  10.1007/JHEP03(2016)105} {\bibfield  {journal} {\bibinfo  {journal} {J. High
  Energy Phys.}\ }\textbf {\bibinfo {volume} {03}},\ \bibinfo {pages} {105}
  (\bibinfo {year} {2016})}\BibitemShut {NoStop}%
\bibitem [{\citenamefont {Can}\ \emph {et~al.}(2015)\citenamefont {Can},
  \citenamefont {Laskin},\ and\ \citenamefont {Wiegmann}}]{can2014geometry}%
  \BibitemOpen
  \bibfield  {author} {\bibinfo {author} {\bibfnamefont {T.}~\bibnamefont
  {Can}}, \bibinfo {author} {\bibfnamefont {M.}~\bibnamefont {Laskin}}, \ and\
  \bibinfo {author} {\bibfnamefont {P.~B}\ \bibnamefont {Wiegmann}},\
  }\bibfield  {title} {\enquote {\bibinfo {title} {{Geometry of Quantum Hall
  States: Gravitational Anomaly and Transport Coefficients}},}\ }\href
  {\doibase 10.1016/j.aop.2015.02.013} {\bibfield  {journal} {\bibinfo
  {journal} {Ann. Phys. (Amsterdam)}\ }\textbf {\bibinfo {volume} {362}},\
  \bibinfo {pages} {752} (\bibinfo {year} {2015})}\BibitemShut {NoStop}%
\bibitem [{\citenamefont {Gromov}\ and\ \citenamefont
  {Abanov}(2014)}]{Gromov-galilean}%
  \BibitemOpen
  \bibfield  {author} {\bibinfo {author} {\bibfnamefont {A.}~\bibnamefont
  {Gromov}}\ and\ \bibinfo {author} {\bibfnamefont {A.~G.}\ \bibnamefont
  {Abanov}},\ }\bibfield  {title} {\enquote {\bibinfo {title}
  {{Density-Curvature Response and Gravitational Anomaly}},}\ }\href {\doibase
  10.1103/PhysRevLett.113.266802} {\bibfield  {journal} {\bibinfo  {journal}
  {Phys. Rev. Lett.}\ }\textbf {\bibinfo {volume} {113}},\ \bibinfo {pages}
  {266802} (\bibinfo {year} {2014})}\BibitemShut {NoStop}%
\bibitem [{\citenamefont {Nguyen}\ \emph {et~al.}(2017)\citenamefont {Nguyen},
  \citenamefont {Can},\ and\ \citenamefont {Gromov}}]{Nguyen-PHD}%
  \BibitemOpen
  \bibfield  {author} {\bibinfo {author} {\bibfnamefont {D.~X.}\ \bibnamefont
  {Nguyen}}, \bibinfo {author} {\bibfnamefont {T.}~\bibnamefont {Can}}, \ and\
  \bibinfo {author} {\bibfnamefont {A.}~\bibnamefont {Gromov}},\ }\bibfield
  {title} {\enquote {\bibinfo {title} {{Particle-Hole Duality in the Lowest
  Landau Level}},}\ }\href {\doibase 10.1103/PhysRevLett.118.206602} {\bibfield
   {journal} {\bibinfo  {journal} {Phys. Rev. Lett.}\ }\textbf {\bibinfo
  {volume} {118}},\ \bibinfo {pages} {206602} (\bibinfo {year}
  {2017})}\BibitemShut {NoStop}%
\bibitem [{Note5()}]{Note5}%
  \BibitemOpen
  \bibinfo {note} {To be really consistent, we have to include higher-spin
  fields in the theory. These fields will also enter the definition of the
  density. Instead of doing this, we take a more phenomenological approach; we
  assume that higher-spin fields have already been integrated out and have
  generated the correction to the density, proportional to $\Delta \protect
  \mathaccentV {hat}05ER$.}\BibitemShut {Stop}%
\bibitem [{\citenamefont {Girvin}(1984)}]{girvin1984particle}%
  \BibitemOpen
  \bibfield  {author} {\bibinfo {author} {\bibfnamefont {S.~M.}\ \bibnamefont
  {Girvin}},\ }\bibfield  {title} {\enquote {\bibinfo {title} {{Particle-Hole
  Symmetry in the Anomalous Quantum Hall Effect}},}\ }\href {\doibase
  10.1103/PhysRevB.29.6012} {\bibfield  {journal} {\bibinfo  {journal} {Phys.
  Rev. B}\ }\textbf {\bibinfo {volume} {29}},\ \bibinfo {pages} {6012}
  (\bibinfo {year} {1984})}\BibitemShut {NoStop}%
\bibitem [{\citenamefont {Levin}\ and\ \citenamefont {Son}(2017)}]{SL}%
  \BibitemOpen
  \bibfield  {author} {\bibinfo {author} {\bibfnamefont {M.}~\bibnamefont
  {Levin}}\ and\ \bibinfo {author} {\bibfnamefont {D.~T.}\ \bibnamefont
  {Son}},\ }\bibfield  {title} {\enquote {\bibinfo {title} {{Particle-Hole
  Symmetry and Electromagnetic Response of a Half-Filled Landau Level}},}\
  }\href {\doibase 10.1103/PhysRevB.95.125120} {\bibfield  {journal} {\bibinfo
  {journal} {Phys. Rev. B}\ }\textbf {\bibinfo {volume} {95}},\ \bibinfo
  {pages} {125120} (\bibinfo {year} {2017})}\BibitemShut {NoStop}%
\bibitem [{\citenamefont {Chui}(1986)}]{chui1986shear}%
  \BibitemOpen
  \bibfield  {author} {\bibinfo {author} {\bibfnamefont {S.~T.}\ \bibnamefont
  {Chui}},\ }\bibfield  {title} {\enquote {\bibinfo {title} {{Shear Modulus of
  Laughlin-Type wave Functions}},}\ }\href {\doibase 10.1103/PhysRevB.34.1409}
  {\bibfield  {journal} {\bibinfo  {journal} {Phys. Rev. B}\ }\textbf {\bibinfo
  {volume} {34}},\ \bibinfo {pages} {1409} (\bibinfo {year}
  {1986})}\BibitemShut {NoStop}%
\bibitem [{\citenamefont {Park}\ and\ \citenamefont
  {Haldane}(2014)}]{park2014guiding}%
  \BibitemOpen
  \bibfield  {author} {\bibinfo {author} {\bibfnamefont {Y.}~\bibnamefont
  {Park}}\ and\ \bibinfo {author} {\bibfnamefont {F.~D.~M.}\ \bibnamefont
  {Haldane}},\ }\bibfield  {title} {\enquote {\bibinfo {title} {{Guiding-Center
  Hall Viscosity and Intrinsic Dipole Moment Along Edges of Incompressible
  Fractional Quantum Hall Fluids}},}\ }\href {\doibase
  10.1103/PhysRevB.90.045123} {\bibfield  {journal} {\bibinfo  {journal} {Phys.
  Rev. B}\ }\textbf {\bibinfo {volume} {90}},\ \bibinfo {pages} {045123}
  (\bibinfo {year} {2014})}\BibitemShut {NoStop}%
\bibitem [{\citenamefont {Schine}\ \emph {et~al.}(2016)\citenamefont {Schine},
  \citenamefont {Ryou}, \citenamefont {Gromov}, \citenamefont {Sommer},\ and\
  \citenamefont {Simon}}]{schine2015synthetic}%
  \BibitemOpen
  \bibfield  {author} {\bibinfo {author} {\bibfnamefont {N.}~\bibnamefont
  {Schine}}, \bibinfo {author} {\bibfnamefont {A.}~\bibnamefont {Ryou}},
  \bibinfo {author} {\bibfnamefont {A.}~\bibnamefont {Gromov}}, \bibinfo
  {author} {\bibfnamefont {A.}~\bibnamefont {Sommer}}, \ and\ \bibinfo {author}
  {\bibfnamefont {J.}~\bibnamefont {Simon}},\ }\bibfield  {title} {\enquote
  {\bibinfo {title} {{Synthetic Landau Levels for Photons}},}\ }\href {\doibase
  10.1038/nature17943} {\bibfield  {journal} {\bibinfo  {journal} {Nature
  (London)}\ }\textbf {\bibinfo {volume} {534}},\ \bibinfo {pages} {671}
  (\bibinfo {year} {2016})}\BibitemShut {NoStop}%
\bibitem [{\citenamefont {Can}\ \emph {et~al.}(2016)\citenamefont {Can},
  \citenamefont {Chiu}, \citenamefont {Laskin},\ and\ \citenamefont
  {Wiegmann}}]{PhysRevLett.117.266803}%
  \BibitemOpen
  \bibfield  {author} {\bibinfo {author} {\bibfnamefont {T.}~\bibnamefont
  {Can}}, \bibinfo {author} {\bibfnamefont {Y.~H.}\ \bibnamefont {Chiu}},
  \bibinfo {author} {\bibfnamefont {M.}~\bibnamefont {Laskin}}, \ and\ \bibinfo
  {author} {\bibfnamefont {P.}~\bibnamefont {Wiegmann}},\ }\bibfield  {title}
  {\enquote {\bibinfo {title} {{Emergent Conformal Symmetry and Geometric
  Transport Properties of Quantum Hall States on Singular Surfaces}},}\ }\href
  {\doibase 10.1103/PhysRevLett.117.266803} {\bibfield  {journal} {\bibinfo
  {journal} {Phys. Rev. Lett.}\ }\textbf {\bibinfo {volume} {117}},\ \bibinfo
  {pages} {266803} (\bibinfo {year} {2016})}\BibitemShut {NoStop}%
\bibitem [{\citenamefont {Klevtsov}(2017)}]{klevtsov2016lowest}%
  \BibitemOpen
  \bibfield  {author} {\bibinfo {author} {\bibfnamefont {S.}~\bibnamefont
  {Klevtsov}},\ }\bibfield  {title} {\enquote {\bibinfo {title} {{Lowest Landau
  Level on a Cone and Zeta Determinants}},}\ }\href {\doibase
  10.1088/1751-8121/aa6e0a} {\bibfield  {journal} {\bibinfo  {journal} {J.
  Phys. A}\ }\textbf {\bibinfo {volume} {50}},\ \bibinfo {pages} {234003}
  (\bibinfo {year} {2017})}\BibitemShut {NoStop}%
\bibitem [{\citenamefont {Can}(2017)}]{can2016central}%
  \BibitemOpen
  \bibfield  {author} {\bibinfo {author} {\bibfnamefont {T.}~\bibnamefont
  {Can}},\ }\bibfield  {title} {\enquote {\bibinfo {title} {{Central Charge
  from Adiabatic Transport of Cusp Singularities in the Quantum Hall
  Effect}},}\ }\href {\doibase 10.1088/1751-8121/aa640e} {\bibfield  {journal}
  {\bibinfo  {journal} {J. Phys. A}\ }\textbf {\bibinfo {volume} {50}},\
  \bibinfo {pages} {174004} (\bibinfo {year} {2017})}\BibitemShut {NoStop}%
\bibitem [{\citenamefont {Gromov}(2016)}]{gromov2016geometric}%
  \BibitemOpen
  \bibfield  {author} {\bibinfo {author} {\bibfnamefont {A.}~\bibnamefont
  {Gromov}},\ }\bibfield  {title} {\enquote {\bibinfo {title} {{Geometric
  Defects in Quantum Hall States}},}\ }\href {\doibase
  10.1103/PhysRevB.94.085116} {\bibfield  {journal} {\bibinfo  {journal} {Phys.
  Rev. B}\ }\textbf {\bibinfo {volume} {94}},\ \bibinfo {pages} {085116}
  (\bibinfo {year} {2016})}\BibitemShut {NoStop}%
\bibitem [{\citenamefont {Golkar}\ \emph
  {et~al.}(2016{\natexlab{b}})\citenamefont {Golkar}, \citenamefont {Nguyen},
  \citenamefont {Roberts},\ and\ \citenamefont {Son}}]{golkar2016higher}%
  \BibitemOpen
  \bibfield  {author} {\bibinfo {author} {\bibfnamefont {S.}~\bibnamefont
  {Golkar}}, \bibinfo {author} {\bibfnamefont {D.~X.}\ \bibnamefont {Nguyen}},
  \bibinfo {author} {\bibfnamefont {M.~M.}\ \bibnamefont {Roberts}}, \ and\
  \bibinfo {author} {\bibfnamefont {D.~T.}\ \bibnamefont {Son}},\ }\bibfield
  {title} {\enquote {\bibinfo {title} {{Higher-Spin Theory of the
  Magnetorotons}},}\ }\href {\doibase 10.1103/PhysRevLett.117.216403}
  {\bibfield  {journal} {\bibinfo  {journal} {Phys. Rev. Lett.}\ }\textbf
  {\bibinfo {volume} {117}},\ \bibinfo {pages} {216403} (\bibinfo {year}
  {2016}{\natexlab{b}})}\BibitemShut {NoStop}%
\bibitem [{\citenamefont {Laskin}\ \emph {et~al.}(2015)\citenamefont {Laskin},
  \citenamefont {Can},\ and\ \citenamefont {Wiegmann}}]{laskin2015collective}%
  \BibitemOpen
  \bibfield  {author} {\bibinfo {author} {\bibfnamefont {M.}~\bibnamefont
  {Laskin}}, \bibinfo {author} {\bibfnamefont {T.}~\bibnamefont {Can}}, \ and\
  \bibinfo {author} {\bibfnamefont {P.}~\bibnamefont {Wiegmann}},\ }\bibfield
  {title} {\enquote {\bibinfo {title} {{Collective Field Theory for Quantum
  Hall States}},}\ }\href {\doibase 10.1103/PhysRevB.92.235141} {\bibfield
  {journal} {\bibinfo  {journal} {Phys. Rev. B}\ }\textbf {\bibinfo {volume}
  {92}},\ \bibinfo {pages} {235141} (\bibinfo {year} {2015})}\BibitemShut
  {NoStop}%
\bibitem [{\citenamefont {Read}\ and\ \citenamefont {Rezayi}(1999)}]{Read1999}%
  \BibitemOpen
  \bibfield  {author} {\bibinfo {author} {\bibfnamefont {N.}~\bibnamefont
  {Read}}\ and\ \bibinfo {author} {\bibfnamefont {E.}~\bibnamefont {Rezayi}},\
  }\bibfield  {title} {\enquote {\bibinfo {title} {{Beyond Paired Quantum Hall
  States: Parafermions and Incompressible States in the First Excited Landau
  Level}},}\ }\href {\doibase 10.1103/PhysRevB.59.8084} {\bibfield  {journal}
  {\bibinfo  {journal} {Phys. Rev. B}\ }\textbf {\bibinfo {volume} {59}},\
  \bibinfo {pages} {8084} (\bibinfo {year} {1999})}\BibitemShut {NoStop}%
\bibitem [{\citenamefont {Jain}(1989)}]{jain1989composite}%
  \BibitemOpen
  \bibfield  {author} {\bibinfo {author} {\bibfnamefont {J.~K.}\ \bibnamefont
  {Jain}},\ }\bibfield  {title} {\enquote {\bibinfo {title} {{Composite-Fermion
  Approach for the Fractional Quantum Hall Effect}},}\ }\href {\doibase
  10.1103/PhysRevLett.63.199} {\bibfield  {journal} {\bibinfo  {journal} {Phys.
  Rev. Lett.}\ }\textbf {\bibinfo {volume} {63}},\ \bibinfo {pages} {199}
  (\bibinfo {year} {1989})}\BibitemShut {NoStop}%
\end{thebibliography}%

\end{document}